\documentclass[letter,11pt]{JHEP3}
\usepackage{graphicx}
\usepackage{psfrag}
\usepackage{fancybox}
\usepackage{ifthen}
\usepackage{amsmath}
\usepackage[errorshow]{tracefnt}
\usepackage{multirow}
\newcommand{\eq}[1]{\begin{equation}#1\end{equation}}
\newcommand{\al}[1]{\begin{align}#1\end{align}}
\newcommand{\subeq}[1]{\begin{subequations}#1\end{subequations}}
\newcommand{\spl}[1]{\begin{split}#1\end{split}}

\newcommand{\arXividhepth}[1]{\href{http://arxiv.org/abs/#1}{\tt arXiv:#1}}

\def\Re           {{\rm Re\hskip0.1em}}
\def\Im           {{\rm Im\hskip0.1em}}

\newcommand\benu{\begin{enumerate}}
\newcommand\eenu{\end{enumerate}}
\newcommand\bit{\begin{itemize}}
\newcommand\eit{\end{itemize}}

\newcommand\g{\gamma}

\newcommand\e{\epsilon}

\newcommand\G{\Gamma}

\newcommand{\boxedeq}[1]{
\begin{equation}
\fbox{
\rule[0.7cm]{0pt}{0pt}
$#1$
\rule[-0.45cm]{0pt}{0pt}
}
\end{equation}
}

\title{Type IIA AdS$_4$ compactifications on cosets, \\
interpolations and domain walls}

\author{Paul Koerber${}^{\diamondsuit}$, Dieter L\"{u}st${}^{\diamondsuit\clubsuit}$
 and Dimitrios Tsimpis${}^{\clubsuit}$ \\

  \begin{itemize}

\item  Max-Planck-Institut f\"{u}r Physik\\
F\"{o}hringer Ring 6, 80805 M\"{u}nchen, Germany
  
\item  Arnold-Sommerfeld-Center for Theoretical Physics\\
Department f\"{u}r Physik, Ludwig-Maximilians-Universit\"{a}t M\"{u}nchen\\
Theresienstra\ss e 37, 80333 M\"{u}nchen, Germany
  \end{itemize}

\bigskip
 E-mail:
\email{koerber@mppmu.mpg.de}, \email{dieter.luest@lmu.de} \& \email{luest@mppmu.mpg.de}, \email{tsimpis@theorie.physik.uni-muenchen.de }}

\abstract{We present a classification of a large class of type IIA
$\mathcal{N}=1$ supersymmetric compactifications to AdS$_4$, based on left-invariant SU(3)-structures
on coset spaces. In the absence of sources
the parameter spaces of all cosets leading to a solution contain regions
corresponding to nearly-K\"{a}hler structure. I.e.\ all these cosets
 can be viewed as deformations of nearly-K\"{a}hler manifolds.
Allowing for (smeared) six-brane/orientifold sources we obtain more possibilities.
In the second part of the paper, we use a simple ansatz, which can
be applied to all six-dimensional coset manifolds considered here, to construct explicit thick domain
wall solutions separating two AdS$_4$ vacua of different radii. We also
consider smooth interpolations between AdS$_4\times \mathcal{M}_6$
and $\mathbb{R}^{1,2}\times \mathcal{M}_7$, where $\mathcal{M}_6$ is a nearly-K\"{a}hler
manifold and $\mathcal{M}_7$ is the G$_2$-holonomy cone over $\mathcal{M}_6$.
}

\preprint{arXiv:0804.0614 [hep-th]\\MPP-2008-29\\LMU-ASC 19/08}
\begin{document}

\section{Introduction and summary}

In recent years,
it has become clear that  compactifications of string theory in the presence of fluxes can
be usefully described in the language of G-structures \cite{gstruc}.
In particular the requirement of $\mathcal{N}=1$ supersymmetry in four dimensions in type II
 for six-dimensional compactification manifolds of SU(3) structure can be conveniently summarized as a set of necessary conditions on the torsion classes of
these manifold \cite{getala} (see \cite{granareview} for a review and further references).
It was subsequently realized that generalized geometry \cite{gengeom} provides a natural framework for the most general $\mathcal{N}=1$ supersymmetric
ansatz in type II, also known as SU(3)$\times$SU(3)-structure, and it was shown in \cite{granaN1} that the supersymmetry conditions can be succinctly
rewritten as differential conditions on a pair of polyforms.

A systematic search for concrete examples of six-dimensional manifolds, suitable for
$\mathcal{N}=1$ compactification to four-dimensional Minkowski space,
has yielded very few examples \cite{granascan}. Moreover, due to a no-go theorem \cite{nogo} these examples require
the presence of orientifold planes, typically smeared. In certain cases, it can be argued that
the latter arise as the large-volume supergravity approximation of bona-fide
string-theory orientifolds.

\begin{table}[h]
\begin{center}
\begin{tabular}{|c|c|}
\hline
$G$ & $H$\\
\hline
\hline
G$_2$ & SU(3)\\
\hline
SU(3)$\times$SU(2)$^2$ & SU(3)\\
\hline
\hline
Sp(2) & S(U(2)$\times$U(1))\\
\hline
SU(3)$\times$U(1)$^2$ & S(U(2)$\times$U(1))\\
\hline
SU(2)$^3\times$U(1) & S(U(2)$\times$U(1))\\
\hline
\hline
SU(3) & U(1)$\times$U(1)\\
\hline
SU(2)$^2\times$U(1)$^2$ & U(1)$\times$U(1)\\
\hline
\hline
SU(3)$\times$U(1) & SU(2)\\
\hline
SU(2)$^3$ & SU(2)\\
\hline
\hline
SU(2)$^2\times$U(1) & U(1)\\
\hline
\hline
SU(2)$^2$ & 1\\
\hline
\end{tabular}
\caption{\label{taba} All six-dimensional manifolds of the type $M=G/H$, where $H$ is a subgroup of SU(3).}
\end{center}
\end{table}

The situation is somewhat better in $\mathcal{N}=1$ compactifications to
four-dimensional anti-de Sitter space \cite{behr,ltads} where the no-go theorem can
be circumvented. For instance, the six-dimensional compact nearly-K\"{a}hler manifolds constitute
a viable starting point for supersymmetric compactifications, without the need
for orientifolds. Recently, it was pointed out in \cite{font} that the  Hopf reductions of
eleven-dimensional
supergravity considered by Nilsson and Pope \cite{np} lead to supersymmetric IIA compactifications that are not nearly-K\"{a}hler, in that the torsion
class ${\cal W}_2$ is non-zero. Necessarily, however, these solutions have vanishing Romans mass. Subsequently,
using twistor-space techniques, the author of
\cite{tomtwistor} constructed compactifications interpolating between the nearly-K\"ahler and vanishing-Romans-mass
cases on two special coset manifolds -- each of which can be seen as a twistor bundle\footnote{Publication \cite{palt} considers the compactification of IIA supergravity
on the coset SU(3)/U(1)$\times$U(1), but without any analysis of the Bianchi identities
of the form-fields.}.

In the present paper we  provide a classification of
a large class of concrete examples of six-dimensional compact manifolds that satisfy
the necessary and sufficient conditions for $\mathcal{N}=1$ compactification with a strict SU(3)-structure
ansatz to four-dimensional anti-de Sitter space. Namely, we consider
compactifications on manifolds
of the type $M=G/H$, where $G$ is a Lie group (not necessarily simple)
and $H$ is a closed subgroup, such that the
action of $G$ on $M$ is effective. Coset spaces were studied some time ago, in the context of the
 Kaluza-Klein approach to unification. For a review see
\cite{zoup} and references therein. For early work, in the context of heterotic string theory,
see \cite{cosetheterotic,govi}; for some recent results see \cite{mpz}.

The requirement of four-dimensional
supersymmetry imposes the condition that the structure group of $T(M)$, the tangent bundle of
the six-dimensional internal manifold $M$, is reduced to SU(3). As we show in appendix \ref{structuregroup},
this translates into the requirement that $H$ be isomorphic to SU(3) or a subgroup thereof.
All possible six-dimensional manifolds $M$ of this type can be easily classified, and consist of the
ones listed in table \ref{taba},
as well as those obtained from the above by replacing any number of SU(2) factors in $G$ by factors
of U(1)$^3$.

As we review in section \ref{ads4solutions}, the necessary and sufficient conditions
for $\mathcal{N}=1$ compactification to  four-dimensional  anti-de Sitter space on manifolds of
SU(3)-structure can be compactly summarized as a set of conditions
on the torsion classes of the internal six-dimensional manifold;
the resulting geometry is then determined by the fluxes \cite{ltads}. In particular, the intrinsic
torsion, $\tau$, of the six-dimensional manifold must be contained in the first two
torsion classes $\mathcal{W}^-_{1,2}$.
In the special case where the second torsion class vanishes, $\mathcal{W}^-_2=0$, the manifold
is called {\it nearly-K\"{a}hler}.

In the absence of sources, there are additional
constraints on the torsion classes: a) the exterior derivative of the second torsion class must be proportional
to the real part of the three-form of the SU(3)-structure, and b) the norm of the
first torsion class is  bounded below by the norm of the second torsion class.
All the
conditions are summarized in table \ref{tabb}.
Note, however,  that in the presence of sources the last two conditions can be relaxed, as we review in the following.
\begin{table}[h]
\begin{center}
\begin{tabular}{|c|}
\hline
$\tau\in \mathcal{W}^-_1\oplus\mathcal{W}^-_2$\\
 $d\mathcal{W}^-_2\propto \Re\Omega$\\
 $3|{\cal W}^-_1|^2\geq |{\cal W}^-_2|^2$ \\
\hline
\end{tabular}
\caption{\label{tabb} Necessary and sufficient conditions on the internal six-dimensional SU(3)-structure manifold
for $\mathcal{N}=1$ compactification to  four-dimensional anti-de Sitter space, in the absence of sources.}
\end{center}
\end{table}

Given the list of table \ref{taba}, one can systematically
search for those manifolds that satisfy the necessary and sufficient conditions for
$\mathcal{N}=1$ compactification to  four-dimensional anti-de Sitter space, listed in table \ref{tabb}.
As we review in  section \ref{cosetreview},
the coset structure of the manifolds is essential for the analysis,
because it allows for the definition of left-invariant one-forms on which the action of the
exterior derivative is completely determined by the structure constants of the
coset. If one further imposes (as we do here) that the SU(3)-structure be {\it left-invariant}\footnote{The
restriction to left-invariant SU(3)-structures is made here in order
to render the problem tractable. We leave the investigation of more general
possibilities for future work.},
the torsion classes of the coset (which can be obtained from the SU(3)-structure by exterior
differentiation) are completely determined in terms of the structure constants. It then
suffices to write down the most general left-invariant ansatz for the SU(3)-structure and
impose that the torsion classes  satisfy the necessary and sufficient conditions
of table \ref{tabb}.
\begin{table}[h]
\begin{center}
\begin{tabular}{|c||c|c|c|c|}
\hline
& SU(2)$\times$SU(2) & \rule[1.2em]{0pt}{0pt} $\frac{\text{SU(3)}}{\text{U(1)}\times \text{U(1)}}$ & $\frac{\text{Sp(2)}}{\text{S}(\text{U(2)}\times \text{U(1)})}$ &$\frac{\text{G}_2}{\text{SU(3)}}$\\
\hline
\hline
\# of parameters  & 1 & 3 & 2 & 1\\
\hline
$\mathcal{W}^-_2\neq 0$ &
No & Yes & Yes & No \\
\hline
\end{tabular}
\caption{\label{tab2} Six-dimensional cosets that satisfy the necessary and sufficient conditions
for $\mathcal{N}=1$ compactification to  four-dimensional anti-de Sitter space, in the absence of sources.}
\end{center}
\end{table}

One then ends up with exactly four possibilities, which are listed in table \ref{tab2}.
The number of arbitrary parameters (moduli) of each solution is indicated in the first row.
More precisely: this is the number of moduli of {\it left-invariant} SU(3)-structures,
such that the conditions of table \ref{tabb} are satisfied.
There is always at least one modulus, corresponding to the overall volume rescaling. Note that
although these moduli can be continuous parameters from the point-of-view of classical supergravity, they are determined in terms of the fluxes  of the solution (as will be explained in more detail in the following section). Since the fluxes are quantized in the full quantum theory, the `moduli' can only assume discrete values.

\begin{figure}
\centering
\psfrag{M=G/H}{$M=G/H$}
\psfrag{M}{$\mathfrak{M}$}
\psfrag{a=c}{$a=c$}
\includegraphics[width=8cm]{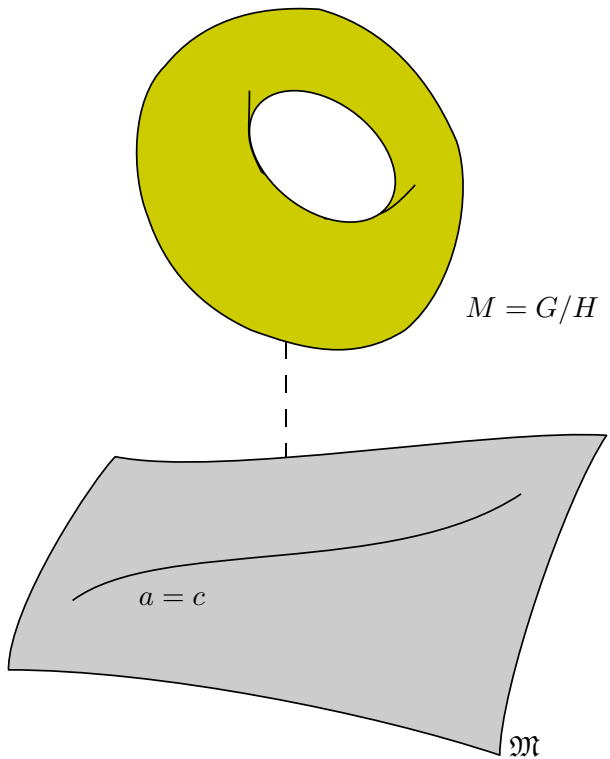}
\caption{The coset space $\frac{\text{Sp(2)}}{\text{S}(\text{U(2)}\times \text{U(1)})}$ fibered over its two-dimensional moduli space
$\mathfrak{M}$. The nearly-K\"{a}hler limit
corresponds to the line $a=c$ in $\mathfrak{M}$, see section 4.3 below. In the full
quantum theory the moduli can only assume discrete values.}
\label{nkmod}
\end{figure}

All cosets of
table \ref{tab2} admit points (more precisely: lines) in their moduli spaces which correspond to
nearly-K\"{a}hler structure (see figure \ref{nkmod}).
Whenever
the moduli space is one-dimensional, i.e.\ whenever the only modulus is the overall volume,
the solution only admits a nearly-K\"{a}hler structure.  In fact, the list of table \ref{tab2} is identical
to the list of all six-dimensional compact homogeneous manifolds that admit a strictly nearly-K\"{a}hler structure \cite{nkmanifolds}\footnote{They are also precisely those coset spaces which were singled out in the first paper in \cite{cosetheterotic}.}.
On the other hand, whenever there are more parameters than
just the volume modulus, i.e.\ whenever the dimension of moduli space is
two or higher, the solution can be deformed away from the nearly-K\"{a}hler line.

The second row (labelled by $\mathcal{W}^-_2\neq 0$) indicates whether or not the coset
admits a left-invariant SU(3)-structure that is {\em not} nearly-K\"{a}hler. This is indeed the case
for the cosets $\frac{\text{SU(3)}}{\text{U(1)}\times \text{U(1)}}$ and $\frac{\text{Sp(2)}}{\text{S}(\text{U(2)}\times \text{U(1)})}$, but not for the
cosets $\text{SU(2)}\times \text{SU(2)}$, $\frac{\text{G}_2}{\text{SU(3)}}$, which only admit a rigid nearly-K\"{a}hler structure.
We stress again that {\it all} cosets of table \ref{tab2} admit nearly-K\"{a}hler structures. In other words,
if a coset admits a structure with $\mathcal{W}^-_2\neq 0$ it also admits a structure with $\mathcal{W}^-_2= 0$,
but not vice versa.

A number of cosets not listed in table \ref{tab2} admit solutions which turn out to be equivalent
to the ones already listed in the table. More precisely, the cosets $\frac{\text{SU(2)}^2\times \text{U(1)}}{\text{U(1)}}$,
$\frac{\text{SU(2)}^3}{\text{SU(2)}}$, $\frac{\text{SU(3)}\times \text{SU(2)}^2}{\text{SU(3)}}$, admit structure constants and
left-invariant SU(3) structures which
turn out to be equivalent to the ones of the SU(2)$\times$SU(2) coset. More details,
as well as the structure constants for each coset and the SU(3)-structure for each
solution, are given in section \ref{cbc}.

All the cosets listed in table \ref{tab2} also admit {\it smeared} six-brane/orientifold sources whose Poincar\'e
dual $j^6$ is proportional
to the real part of the three-form of the SU(3)-structure: $j^6\propto \Re\Omega$.
If one allows for smeared six-brane/orientifold sources that violate this proportionality condition,
then there is one additional possibility: $\frac{\text{SU(3)}\times \text{U(1)}}{\text{SU(2)}}$, with the topology of $S^5\times S^1$.
Table \ref{cab} lists the cosets that satisfy the necessary and sufficient conditions
for $\mathcal{N}=1$ compactification to four-dimensional anti-de Sitter space, in the presence of smeared sources.
The third row indicates whether or not the Poincar\'e dual of the source is proportional to $\Re\Omega$.
In the case of SU(2)$\times$SU(2) there are solutions both with $j^6$ proportional, and not
proportional to $\Re\Omega$.
\begin{table}
\begin{center}
\begin{tabular}{|c||cc|c|c|c|c|}
\hline
& \multicolumn{2}{|c|}{SU(2)$\times$SU(2)} & \rule[1.2em]{0pt}{0pt} $\frac{\text{SU(3)}}{\text{U(1)}\times \text{U(1)}}$ & $\frac{\text{Sp(2)}}{\text{S}(\text{U(2)}\times \text{U(1)})}$ &$\frac{\text{G}_2}{\text{SU(3)}}$
& $\frac{\text{SU(3)}\times \text{U(1)}}{\text{SU(2)}}$\\
\hline
\hline
\# of parameters  & \hspace{0.25cm} 2 \hspace{0.25cm} &  4  & 4 & 3 & 2 & 4\\
\hline
$\mathcal{W}^-_2\neq 0$ &
No & Yes & Yes & Yes & No & Yes \\
\hline
$j^6\propto \Re\Omega$  & Yes & No & Yes & Yes & Yes & No\\
\hline
\end{tabular}
\caption{\label{cab} Six-dimensional cosets that satisfy the necessary and sufficient conditions
for $\mathcal{N}=1$ compactification to  four-dimensional anti-de Sitter space, in the presence of smeared
six-brane/orientifold sources. In our parameter counting we now also include the number of sources,
which is of course again a discrete quantity. For SU(2)$\times$SU(2) we distinguish two cases depending on whether or not the source term is
proportional to $\Re \Omega$.}
\end{center}
\end{table}

To conclude the discussion of the coset vacua
let us make a remark on possible type IIA/IIB AdS$_4$
supersymmetric backgrounds within the class of coset geometries with more general G-structure than
strict SU(3).\footnote{For more details on the meaning of ``strict SU(3)'', ``static SU(2)'' and ``SU(3)$\times$SU(3)'' see
\cite{granaN1,kt}.}  Obviously for static SU(2), but also for
SU(3)$\times$SU(3)-structure if one insists on left-invariant structures, supersymmetry
requires $H$ to be a subgroup of SU(2).
This leaves only the last four entries of table \ref{taba} as candidates.
We have only found a
static SU(2) IIB solution\footnote{Recall that it is impossible to have
supersymmetric IIA AdS$_4$ solutions with static SU(2) structure \cite{bovy}.}
on $\frac{\text{SU(3)}\times\text{U(1)}}{\text{SU(2)}}$, which is T-dual to
the IIA SU(3) solution on the same coset, and
one on $\frac{\text{SU(2)}^2\times\text{U(1)}}{\text{U(1)}}$, which
is T-dual to the IIA SU(3) solution on SU(2)$\times$SU(2).

In the final part of the paper,
using a simple ansatz, which can
be applied to all six-dimensional coset manifolds $\mathcal{M}_6=M$ considered here,
we have been able to obtain smooth interpolations  between two AdS$_4$ vacua of different radii.
These solutions can be interpreted as domain walls in the four noncompact dimensions,
and they necessarily contain `thick' branes. By that we mean branes whose profile in the radial direction (the direction
transverse to the wall) is not a delta-function, but is nevertheless
localized -- in the sense that it falls off to zero far from the wall.
However, we have been unable to obtain explicit profiles
of non-pathological smooth
interpolations between AdS$_4\times \mathcal{M}_6$
and $\mathbb{R}^{1,2}\times \mathcal{M}_7$,
where $\mathcal{M}_7$ is the Hitchin lift of $\mathcal{M}_6$.

\section{Review of AdS$_4$ solutions}\label{ads4solutions}

The most general form of $\mathcal{N}=1$ compactifications of IIA supergravity to AdS$_4$ with the
ansatz $\eta^{(1)} \propto \eta^{(2)}$ for the internal supersymmetry generators
(the strict SU(3)-structure ansatz) was given by two of the present authors in \cite{ltads}. These vacua must have
constant warp factor and dilaton. Setting the warp factor to one, the solutions of \cite{ltads} are given by\footnote{\label{convfootn}As opposed to \cite{ltads} we do not use superspace conventions. Furthermore we use here the string frame and
put $m=-2 m_{\text{there}}, H=-H_{\text{there}}, J=-J_{\text{there}}, F_2=-2 m_{\text{there}} B'$ and $F_4=-G$.}:
\begin{subequations}
\label{ltsol}
\begin{align}
H &=\frac{2m}{5} e^{\Phi}\Re \Omega \, ,\\
F_2&=\frac{f}{9}J+F'_2 \, , \\
F_4&=f\mathrm{vol}_4+\frac{3m}{10} J\wedge J \, , \\
W e^{i \phi} &=-\frac{1}{5} e^{\Phi}m+\frac{i}{3} e^{\Phi }f \, .
\end{align}
\end{subequations}
In the above ($J$, $\Omega$) is the SU(3)-structure of the internal six-manifold,
i.e.\ $J$ is a real two-form and $\Omega$ is a complex three form such that:
\subeq{\al{\label{suthree}
\Omega\wedge J&=0 \, , \\
\label{suthreenorm}
\Omega\wedge\Omega^*&=\frac{4i}{3}J^3\neq 0
~.}}
$f$, $m$ are constants parameterizing the solution: $f$ is the Freund-Rubin parameter, while $m$
is the mass of Romans' supergravity \cite{roma} -- which can be identified with $F_0$ in the `democratic'
formulation \cite{democratic}. $e^{i\phi}$ is a phase associated to the internal supersymmetry generators
$\eta^{(2)}_+ = e^{i \phi} \eta^{(1)}_+$.
$W$ is defined by the following relation for the AdS Killing spinors
\eq{
\label{defW}
\nabla_{\mu} \zeta_- = \frac{1}{2} W \gamma_{\mu} \zeta_+ \, .
}
The radius of AdS$_4$ is given by $|W|^{-1}$. The two-form $F'_2$ is the primitive part of
$F_2$ (i.e.\ it is in the $\bf{8}$ of SU(3)) and is constrained by the Bianchi identity:
\eq{
\mathrm{d} F'_2=(\frac{2}{27}f^2-\frac{2}{5}m^2 )e^{\Phi} \Re \Omega - j^{6} ~,\label{ltsolb}
}
where we have added a source for D6-branes/O6-planes on the right-hand side.
We immediately see that in the absence of sources the second constraint of
table \ref{tabb} holds, i.e.\ $d \mathcal{W}^-_{2}\propto \Re\Omega$. However
in the presence of nonzero $j^6$, this constraint may be relaxed.

The general properties of supersymmetric sources and their consequences for the integrability
of the supersymmetry equations were recently discussed by two
of the present authors in \cite{kt} within the framework of generalized geometry.
It was shown in this reference that, under certain mild assumptions,
supersymmetry guarantees that the appropriately source-modified Einstein equation and dilaton
equation of motion are automatically satisfied if the source-modified Bianchi identities are
satisfied. For this to work the source must be supersymmetric, which means it
must be generalized calibrated as in \cite{gencal}.

Finally, the only nonzero torsion classes of the internal manifold are ${\cal W}^-_1,{\cal W}^-_2$
such that
\subeq{
\label{torsionclasses}
\al{
\label{torsionclass1}
d J&=-\frac{3}{2}i \, \mathcal{W}_1^- \Re \Omega\, , \\
d \Omega&= \mathcal{W}^-_1 J\wedge J+\mathcal{W}^-_2 \wedge J
~.}}
Moreover, they are given by:
\eq{
{\cal W}^-_1=-\frac{4i}{9} e^{\Phi} f  \, , \qquad
{\cal W}^-_2=-i e^{\Phi} F'_2 \,  .
\label{ltsolc}
}
For the following it will be convenient to also
introduce $c_1:= -\frac{3}{2}i \, \mathcal{W}_1^- $, which appears in \eqref{torsionclass1}.
In addition, for vanishing sources or for sources proportional to $\Re \Omega$ we can also define
$c_2$ by
\eq{
\label{c2def}
d \mathcal{W}_2^- = i c_2 \, \Re \Omega \, .
}
One can show \cite{ltads} that
\eq{
\label{c2expr}
c_2 = - \frac{1}{8} |{\cal W}^-_2|^2 \, .
}

It was further noted in \cite{ltads} that, for vanishing $j^{6}$,  the parameters $f$, $m$ of the
solution obey the bound:  $f^2\geq{27}/{5}m^2$, which follows from
$|{\cal W}^-_2|^2\geq0$, \eqref{ltsolb}, \eqref{c2def} and \eqref{c2expr}, with equality for
nearly-K\"ahler manifolds. However, to determine whether a given geometry (${\cal W}_1^-,{\cal W}_2^-$) corresponds
to a vacuum without orientifold sources, the following bound is more relevant
\begin{align}
\frac{16}{5} e^{2\Phi} m^2 = 3|{\cal W}^-_1|^2-|{\cal W}^-_2|^2\geq 0
~,
\label{condition}
\end{align}
where we have defined $|\Phi|^2:= \Phi^*_{mn}\Phi^{mn}$, for any two-form $\Phi$.
Incidentally, let us note that
condition (\ref{condition}) turns out to be too stringent to be satisfied for
any nilmanifold whose only nonzero
torsion classes are ${\cal W}^-_{1,2}$ \cite{ktu}.

Allowing, however, for a nonzero  source, $j^{6}\neq 0$,
effectively relaxes this  constraint. As a particular example let us consider:
\begin{align}
j^{6}=-\frac{2}{5}e^{-\Phi}\mu \, \Re\Omega~,
\label{jo}
\end{align}
where $\mu$ is an arbitrary real parameter, so that $-\mu$  is proportional to
the orientifold/D6-brane tension
($\mu$ is positive for orientifolds and negative for D6-branes).
The addition of this source term  was first considered in \cite{acha}.
Eq.~(\ref{jo}) above guarantees that the calibration conditions, which
for D6-branes/O6-planes read
\eq{
\label{calcond}
j^6 \wedge \Re \Omega = 0 \, , \qquad j^6 \wedge J = 0 \, ,
}
are satisfied and thus the source wraps supersymmetric cycles.
The bound (\ref{condition}) should now
be replaced by:
\begin{align}\label{boundtra}
\mu\geq\frac{5}{16}\left(|{\cal W}^-_2|^2-3|{\cal W}^-_1|^2\right)~.
\end{align}
Since $\mu$ can be taken to be arbitrary the above equation
can always be satisfied, and therefore no longer imposes any constraint on the
torsion classes of the manifold.

Let us also note that it is possible to consider the inclusion of more general supersymmetric
orientifold six-plane sources, {\em not} given by eq.~(\ref{jo}). In this case the second constraint of
table \ref{tabb}, i.e.\ the constraint $d \mathcal{W}^-_{2}\propto \Re\Omega$,
is relaxed. We will still require this source to satisfy the calibration conditions \eqref{calcond}.

{\it In summary}: In the absence of sources the necessary and sufficient conditions for
$\mathcal{N}=1$ compactifications with strict SU(3)-structure to four-dimensional AdS$_4$ space
are those listed in table \ref{tabb}. However in the presence of sources the last
two of the three constraints may be relaxed. In particular the third constraint can always be
relaxed by the addition of orientifold/D6-brane sources of the form (\ref{jo}).

\section{Coset spaces and left-invariant SU(3)-structures}\label{cosetreview}

In this section we give a brief review of some well-known facts
about coset spaces, with special emphasis on the
material that will be useful to us in the following (for more extensive reviews see~\cite{zoup,cosetrev1,cosetrev2}).

Thanks to the uniqueness theorem quoted in appendix \ref{structuregroup}, in dealing
with coset spaces of the form $G/H$ it suffices to examine the corresponding algebras
$\mathfrak{g}$, $\mathfrak{h}$. Let $\{ \mathcal{H}_a\}$ be a basis of
generators of the algebra $\mathfrak{h}$, and let $\{ \mathcal{K}_i\}$ be a basis of the
complement $\mathfrak{k}$ of $\mathfrak{h}$ inside $\mathfrak{g}$, i.e.\ $a=1,\dots,$ dim($H$) and
$i=1,\dots,$ dim($G$)$-$dim($H$). We define the structure constants as follows:
\eq{\spl{\label{commut}
[\mathcal{H}_a,\mathcal{H}_b]&=f^c{}_{ab}\mathcal{H}_c \, , \\
[\mathcal{H}_a,\mathcal{K}_i]&=f^j{}_{ai}\mathcal{K}_j+f^b{}_{ai}\mathcal{H}_b \, , \\
[\mathcal{K}_i,\mathcal{K}_j]&=f^k{}_{ij}\mathcal{K}_k+f^a{}_{ij}\mathcal{H}_a~.
}}
If $H$ is connected and semisimple, or compact -- as is indeed the case for each $H$
listed in table \ref{taba} -- one can always find a basis of generators $\{ \mathcal{K}_i\}$
such that the structure constants $f^b{}_{ai}$ vanish \cite{cosetrev1,knb}. In other
words: $[\mathcal{H}, \mathcal{K}]\subset \mathcal{K}$, in which case
the coset $G/H$ is called {\it reductive}.

Let $x^m$, $m=1,\dots,$dim($G$)$-$dim($H$), be local coordinates on $G/H$ and let
$L(x)$ be a coset representative. The decomposition of the Lie-algebra valued one form
\al{
L^{-1}dL= e^i\mathcal{K}_i+\omega^a\mathcal{H}_a
~,}
defines a coframe $e^i(x)$ on $G/H$. Moreover, using the commutation relations
(\ref{commut}), we find
\al{\label{dcommut}
de^i=-\frac{1}{2}f^i{}_{jk}e^j\wedge e^k-f^i{}_{aj}\omega^a\wedge e^j
~.}
We are interested in forms that are {\em left-invariant} under the action
of $G$ on $G/H$. One can show that this is the case if and only if for the $p$-form
\eq{
\phi=\frac{1}{p!}\phi_{i_1\dots i_p}e^{i_1}\wedge\dots \wedge e^{i_p} \, ,
}
its components $\phi_{i_1\dots i_p}$ are constants {\em and}
\eq{\label{leftinv}
f^j{}_{a[i_1}\phi_{i_2\dots i_p]j}=0~.
}
If we then take the exterior derivative $d\phi$, condition \eqref{leftinv} ensures
that the part coming from the second term in \eqref{dcommut} drops out so we find that
the exterior derivative preserves the left-invariance property. As an aside one can show that
harmonic forms must be left-invariant and thus the cohomology of the coset manifold
is isomorphic to the cohomology of left-invariant forms.

The strategy we follow in this paper is to restrict ourselves to cosets with left-invariant
SU(3)-structure. In other words, we demand that $(J,\Omega)$ be left-invariant forms
on $G/H$. From eq.~(\ref{dcommut}) it then follows that, given the structure constants of the coset in eq.~(\ref{commut}),
the exterior derivatives $(dJ,d\Omega)$ can be explicitly evaluated. On the other hand,
the first condition of table \ref{tabb} is equivalent to the statement that
$\mathcal{W}^-_{1,2}$ are the only non-vanishing torsion classes of the coset.
As this is not the most general form of
$(dJ,d\Omega)$, this condition
 imposes a constraint on $(J,\Omega)$, which may not have any solutions.
Provided solutions exist, one can immediately read off the torsion classes
$\mathcal{W}^-_{1,2}$. Finally, the second and third conditions of table \ref{tabb}
can be examined to determine whether or not the solutions require the presence of sources.

The procedure described above, when applied to each of the cosets listed in table \ref{taba}, leads
to the results summarized in the introduction. The details of the analysis in each case
are presented in section \ref{cbc}.

\section{Case by case analysis}\label{cbc}

In this section we present the details of the analysis for each coset
listed in table \ref{taba}. As explained in section \ref{cosetreview},
our procedure is as follows: For each coset we first write down the most general
left-invariant ansatz for $(J,\Omega)$. We then impose
 the SU(3)-structure conditions (\ref{suthree}). In addition,
we have to demand that the resulting
metric, implicitly defined by $(J,\Omega)$, be positive.
Next we take into account the structure constants of the coset
in order to evaluate $(dJ,d\Omega)$, using eq.~(\ref{dcommut}). Finally we impose
equations (\ref{torsionclasses}). In case solutions exist, we read off the
torsion classes $\mathcal{W}^-_{1,2}$ and we examine whether or not the Bianchi
identity (\ref{ltsolb}) requires the presence of sources. The results of this analysis
were summarized in the introduction, tables \ref{tab2} and \ref{cab}.

Some further remarks about the presentation in the remainder of this section: in
each case we first give the Betti numbers (which can also be straightforwardly
evaluated) and the structure constants
of the coset. We assume that $\mathfrak{g}$ is generated by $\{E_I\}$, $I=1,\dots, \text{dim}(G)$,
such that
\al{
[E_I,E_J]=f^K{}_{IJ}E_K
~,}
and our labelling is such that the $E_I$ with $I=1,\ldots,6$ correspond to the $\mathcal{K}_i$
(spanning $\mathfrak{k}$) and the $E_I$ with $I=7,\ldots,6+\text{dim}(H)$ correspond to the $\mathcal{H}_a$ (spanning $\mathfrak{h}$).
Then follows the solution (in case it exists)
for the SU(3)-structure $(J,\Omega)$, expressed
in terms of some set of parameters. The conditions on these parameters imposed by the
normalization of $\Omega$ (eq.~(\ref{suthreenorm})) and the positivity
of the metric are listed explicitly. We then give
the explicit form of the torsion classes $\mathcal{W}^-_{1,2}$.

It is always understood, unless otherwise stated, that each solution
satisfies the Bianchi identity (\ref{ltsolb}) in the absence of sources. We therefore
also explicitly list the condition imposed by the bound (\ref{condition}). As
explained in section \ref{ads4solutions},  one can always add O6/D6 sources of the
form (\ref{jo}). The mass parameter $m$ is then no longer determined by \eqref{condition}
and thus becomes an extra free parameter --- also counted in table \ref{cab} --- related to the number
of sources. Whenever there exist solutions with sources
that are not of the form (\ref{jo}), it is stated explicitly.

\subsection{SU(2)$\times$SU(2)}\label{SU2SU2}

\begin{tabular}{|c||c|c|c|}
\hline
\multirow{2}{*}{Betti numbers} & $b_1$ & $b_2$ & $b_3$ \\
\cline{2-4}
& 0 & 0 & 2 \\
\hline
\end{tabular}

The structure constants in this case are
\eq{
f^1{}_{23} = f^4{}_{56}=1 \, , \qquad \mathrm{cyclic}
~.}
In \cite{nkmanifolds} it was shown that there is always a change of basis
preserving the form of the structure constants that brings $J$ in diagonal form
\eq{
J = a e^1 \wedge e^4 + b e^2 \wedge e^5 + c e^3 \wedge e^6 \, .
}
The most general solution to \eqref{ltsolb}, \eqref{torsionclasses}, \eqref{ltsolc} and \eqref{condition} without sources, $j^6=0$, is the nearly-K\"ahler one:
\eq{\label{nksolSU2SU2}\spl{
J & = a(e^{14}+e^{25}+e^{36}) \, , \\
\Omega & = d
 \left( e^{156} + e^{426} + e^{453} - e^{126} - e^{153} - e^{423} \right) \\
& - \frac{2i d}{\sqrt{3}} \left[ e^{123} + e^{456} - \frac{1}{2} \left( e^{156} + e^{426} + e^{453} \right)
- \frac{1}{2} \left( e^{423} + e^{153} + e^{126} \right) \right] \, .
}}
with $a$, the overall scale of the internal geometry, the only free parameter and
\eq{\label{nksolparSU2SU2}\spl{
 a & > 0 \, , \qquad \text{metric positivity} \, , \\
 d^2 & = \frac{2}{\sqrt{3}}a^3 \, , \qquad \text{normalization of } \Omega \, , \\
 c_1 & := -\frac{3i}{2} {\cal W}_1^-= \frac{a}{d} \, , \\
 {\cal W}_2^- & = 0 \, , \\
 e^{2\Phi} m^2 & = \frac{5}{12} c_1^2  \,  .
 }}
%
A different solution is possible with a source not proportional to $\Re \Omega$.
We have then
\eq{\spl{
J & = a e^{14}+ b e^{25}+c e^{36} \, , \\
\Omega & = -\frac{1}{c_1} \Bigg\{a (e^{234} - e^{156}) + b (e^{246} - e^{135}) + c (e^{126} - e^{345}) \\
& -\frac{i}{h} \Big[ -2 \, abc (e^{123}+e^{456}) + a(b^2+c^2-a^2) (e^{234}+e^{156}) + b(a^2+c^2-b^2) (e^{153}+e^{426}) \\
& + c(a^2+b^2-c^2) (e^{345}+e^{126}) \Big] \Bigg\} \, ,
}}
with $a,b$ and $c$ three free parameters and
\eq{\spl{
abc & > 0 \, ,  \qquad \text{metric positivity} \, , \\
h & = \sqrt{2 \, a^2 b^2 + 2 \, b^2 c^2 + 2 \, a^2 c^2 - a^4 - b^4 - c^4} \, , \\
& \text{and thus} \quad 2 \, a^2 b^2 + 2 \, b^2 c^2 + 2 \, a^2 c^2 - a^4 - b^4 - c^4 > 0 \, , \\
c_1^2 & = \frac{h}{2abc} \, , \\
 {\cal W}_2^- & = -\frac{2i}{3 h c_1} \Bigg[\frac{(b^2-c^2)^2 + a^2(-2a^2 + b^2 +c^2)}{bc} e^{14}
 +\frac{(c^2-a^2)^2 + b^2(-2b^2 + c^2 +a^2)}{ac} e^{25}  \\
&  +\frac{(a^2-b^2)^2 + c^2(-2c^2 + a^2 +b^2)}{ab} e^{36}
 \Bigg] \, .
}}
One can check that $d \mathcal{W}^-_2$ is not proportional to $\Re \Omega$ unless $|a|=|b|=|c|$,
which brings us back to the above solution. The source can have total negative or positive tension.
In the latter case this geometry can be created with strictly D-brane sources.


\subsection{$\frac{\text{SU(3)}}{\text{U(1)}\times \text{U(1)}}$}\label{SU3qU1U1}

\begin{tabular}{|c||c|c|c|}
\hline
\multirow{2}{*}{Betti numbers} & $b_1$ & $b_2$ & $b_3$ \\
\cline{2-4}
& 0 & 2 & 0 \\
\hline
\end{tabular}

This space is also known as the flag manifold $\mathbb{F}(1,2;3)$ or
the twistor space $\text{Tw}(\mathbb{CP}^2)$.

We choose a basis such that the structure constants of SU(3) are given by
\eq{
f^1{}_{54}=f^1{}_{36}=f^2{}_{46}=f^2{}_{35}=f^3{}_{47}=f^5{}_{76}=\frac{1}{2} \, , \,
\quad f^1{}_{27}=1 \, , \quad f^3{}_{48}=f^5{}_{68}=\frac{\sqrt{3}}{2} \, ,
~\mathrm{cyclic}~.}
These
can be obtained
from the Gell-Mann structure constants $f_{\text{GM}ijk}$ using the permutation
$(12456738)$. The U(1)$\times$U(1) is then generated by $E^7$ and $E^8$.

The $G$-invariant two-forms and three-forms are spanned by
\eq{
\{e^{12},e^{34},e^{56}\}\, , \qquad \{\rho=e^{245}+e^{135}+e^{146}-e^{236},\hat{\rho}=e^{235}+e^{136}+e^{246}-e^{145}\} \, ,
}
respectively, and there are no invariant one-forms. With the two invariant three-forms, one can construct exactly two invariant
almost complex structures: $J$ associated to $\rho +i \hat{\rho}$ and $-J$ associated to $\rho - i \hat{\rho}$.
Also, with only these two invariant three-forms there is no room for a source not proportional to $\Re \Omega$.

The most general solution is then given by
\eq{\spl{
J & = - a e^{12} + b e^{34} - c e^{56} \, , \\
\Omega & = d \left[ (e^{245}+e^{135}+e^{146}-e^{236}) + i (e^{235}+e^{136}+e^{246}-e^{145}) \right] \, ,
}}
with $a,b$ and $c$ three free parameters and
\eq{\spl{
 a & > 0 , \, b>0, \, c>0 \, , \qquad \text{metric positivity} \, , \\
 d^2 & = abc, \qquad \text{normalization of } \Omega \, , \\
 c_1 & := -\frac{3i}{2} {\cal W}_1^-= -\frac{a+b+c}{2d} \, , \\
 {\cal W}_2^- & = -\frac{2i}{3d} \left[ a(2a-b-c) e^{12} + b (a-2b+c)e^{34} + c(-a-b+2c) e^{56} \right] \, , \\
 c_2 & := - \frac{1}{8} |{\cal W}_2^-|^2 = - \frac{2}{3abc} \left( a^2 + b^2 + c^2 - (ab+ac+bc)\right) \, , \\
 \frac{2}{5} e^{2\Phi} m^2 & = c_2 + \frac{1}{6} c_1^2 = \frac{1}{8abc} \left[- 5(a^2+b^2+c^2) + 6 (ab+ac+bc) \right] \ge 0 \,  .
}}

The nearly-K\"ahler limit corresponds to $a=b=c$.

We can also make the connection with the results of \cite{tomtwistor} by defining the complex one-forms
\eq{
e^{z^1} = a^{1/2} \left( - e^2 + i e^1 \right) \, , \qquad
e^{z^2} = b^{1/2} \left( - e^3 + i e^4 \right) \, , \qquad
e^{z^3} = c^{1/2} \left( - e^6 + i e^5 \right) \, ,
}
which satisfy\footnote{Note that one has now to take into account the second term on the RHS of \eqref{dcommut} as
these complex one-forms are not left-invariant.}
\eq{
d \left( \begin{array}{c} e^{z^1}\\ e^{z^2} \\ e^{z^3} \end{array}\right)=
\left(\begin{array}{cc} - \alpha & 0|_{2 \times 1} \\ 0|_{1 \times 2} & \text{Tr} \, \alpha \end{array}\right)
\left(\begin{array}{c} e^{z^1}\\ e^{z^2} \\ e^{z^3} \end{array}\right)
-\frac{i}{2c^{1/2}}
\left(\begin{array}{c} \left(\frac{a}{b}\right)^{1/2} e^{\bar{z}^2} \wedge e^{\bar{z}^3} \\
\left(\frac{b}{a}\right)^{1/2} e^{\bar{z}^3} \wedge e^{\bar{z}^1} \\
\left(\frac{c}{(ab)^{1/2}}\right) e^{\bar{z}^1} \wedge e^{\bar{z}^2}
\end{array}\right) \, ,
}
with $\alpha$ the anti-hermitian matrix of one forms
\eq{
\alpha = i\left( \begin{array}{cc} \omega^7 & 0 \\ 0 & -\frac{1}{2} \omega^7 - \frac{\sqrt{3}}{2} \omega^8 \end{array}\right) \, .
}
If $a=b$ these equations take (up to conventions) the form of eq.~(3.10) of \cite{tomtwistor}
with $R=-2c^{1/2}$ and $\sigma=c/a$.
By having imposed eq.~(3.10) therein, we see that
the construction of \cite{tomtwistor} misses the possibility $a\neq b$.

\subsection{$\frac{\text{Sp(2)}}{\text{S}(\text{U(2)}\times \text{U(1)})}$}\label{SO5qSU2U1}

\subsubsection*{Maximal embedding}

\begin{tabular}{|c||c|c|c|}
\hline
\multirow{2}{*}{Betti numbers} & $b_1$ & $b_2$ & $b_3$ \\
\cline{2-4}
& 0 & 1 & 0 \\
\hline
\end{tabular}

The algebra sp(2) $\approx$ so(5) is generated by
traceless antisymmetric matrices $\{ J^{(ij)}| ~i,j=1,\dots, 5  \}$ given by
\al{\label{sp2basis}
\left(J^{(ij)}\right)_{kl}=\delta^i_{k}\delta^j_{l}-\delta^i_{l}\delta^j_{k}
~.}
These satisfy the following commutation relations:
\eq{
[J^{(ij)},J^{(kl)}]=\frac{1}{2}\left(\delta^{il}J^{(jk)}+  \delta^{jk}J^{(il)}
-\delta^{jl}J^{(ik)}-\delta^{ik}J^{(jl)}\right)~.
}
The maximal embedding of su(2)$\oplus$u(1) into sp(2) can be realized by taking su(2) $\approx$ so(3) to be generated by
$\{ J^{(12)}, J^{(13)}, J^{(23)} \}$ and u(1) $\approx$ so(2) to be generated by $J^{(45)}$. Let us introduce the following
notation:
\eq{
\{ E_7, E_8, E_9,  E_{10}\}:=\{ J^{(12)}, J^{(13)}, J^{(23)} , J^{(45)}\}~,
} and
\eq{
\{ E_1, \dots,  E_{6}\}:=\{ J^{(14)}, J^{(15)}, J^{(24)}, J^{(25)}, J^{(34)}, J^{(35)} \}~.
}
It follows that in this basis the structure constants are totally antisymmetric, with:
\eq{
f^{7}{}_{89}=f^{7}{}_{13}=f^{7}{}_{24}=f^{8}{}_{15}=f^{8}{}_{26}=f^{9}{}_{35}=f^{9}{}_{46}=f^{10}{}_{12}=f^{10}{}_{34}=f^{10}{}_{56}=-\frac{1}{2}~,
}
being the only nonzero ones.  One can check that $[\mathfrak{k},\mathfrak{k}]=\mathfrak{h}$, as expected
for a symmetric coset space in the canonical decomposition.

While there is an invariant two-form: $e^{12}+e^{34}+e^{56}$,
there are no invariant one- or three-forms, and thus there is no solution.

\subsubsection*{Nonmaximal embedding}

\begin{tabular}{|c||c|c|c|}
\hline
\multirow{2}{*}{Betti numbers} & $b_1$ & $b_2$ & $b_3$ \\
\cline{2-4}
& 0 & 1 & 0 \\
\hline
\end{tabular}

This space is topologically equivalent to
$\mathbb{CP}^3$, which can also be viewed as the twistor space $\text{Tw}(S^4)$.

The nonmaximal embedding is realized by embedding su(2)$\oplus$u(1) into an su(2)$\oplus$su(2) $\approx$ so(4)
subgroup of sp(2). Using the basis (\ref{sp2basis}), let so(4) be the subgroup generated by
$\{ J^{(ij)}| ~i,j=1,\dots, 4  \}$. The isomorphism su(2)$\oplus$su(2) $\approx$ so(4) can be realized explicitly by
noting that the two su(2) subalgebras are generated by $\{ E_{i}| ~i=5, 6, 7  \}$ and
$\{ E_{i}| ~i=8, 9, 10  \}$, where:
\eq{\spl{
 E_{i+4}&:= \frac{1}{2} \varepsilon^{ijk}J^{(jk)}+J^{(i4)}, \\
 E_{i+7}&:=  \frac{1}{2} \varepsilon^{ijk}J^{(jk)}-J^{(i4)}, ~~~~~ i=1,2,3
~.}}
The remaining generators are given by $E_{i}:=\sqrt{2}J^{(i5)} , ~~i=1,\dots, 4$. With the above definitions
the structure constants are totally antisymmetric. The nonzero ones are given by:
\eq{\spl{
& f^5{}_{41}=f^5{}_{32}=f^6{}_{13}=f^6{}_{42}=\frac{1}{2} \, , \qquad f^7{}_{56}=f^{10}{}_{89}=-1 \, , \\
& f^7{}_{21}=f^7{}_{43}=f^8{}_{14}=f^8{}_{32}=f^9{}_{13}=f^9{}_{24}=f^{10}{}_{34}=f^{10}{}_{21}=\frac{1}{2} \, ,
}}
The su(2)$\oplus$u(1) subalgebra is generated by $E_{7},\dots, E_{10}$.

%

The $G$-invariant two-forms and three-forms are spanned by
\eq{
\{e^{12}+e^{34},e^{56}\}\, , \qquad \{\rho=e^{245}-e^{135}-e^{146}-e^{236},\hat{\rho}=e^{235}+e^{246}+e^{145}-e^{136}\} \, ,
}
respectively, and there are no invariant one-forms. The source (if present) must be proportional to $\Re \Omega$.

The most general solution is then given by
\eq{\spl{
J & = a (e^{12} + e^{34}) - c e^{56} \, , \\
\Omega & = d \left[ (e^{245}-e^{236}-e^{146}-e^{135}) + i (e^{246}+e^{235}+e^{145}-e^{136}) \right] \, ,
}}
with $a$ and $c$ two free parameters and
\eq{\spl{
 a & > 0 \, , \quad c>0, \qquad \text{metric positivity} \, , \\
 d^2 & = a^2 c, \qquad \text{normalization of } \Omega \, , \\
 c_1 & := -\frac{3i}{2} {\cal W}_1^-= \frac{2a+c}{2d} \, , \\
 {\cal W}_2^- & = -\frac{2i}{3d} \left[ a(a-c) (e^{12} + e^{34}) + 2c(a-c) e^{56} \right] \, , \\
 c_2 & := - \frac{1}{8} |{\cal W}_2^-|^2 = - \frac{2}{3a^2c}  (a-c)^2 \, , \\
 \frac{2}{5} e^{2\Phi} m^2 & = c_2 + \frac{1}{6} c_1^2 = \frac{1}{8a^2c} \left[- 4a^2-5c^2+12ac \right] \ge 0 \,  .
}}

Note that if we set $a=b$ in the $\frac{\text{SU(3)}}{\text{U(1)}\times \text{U(1)}}$ solution we get the same result as above. The nearly-K\"ahler limit corresponds
to further setting $a=c$.

Again we can make the connection with the results of \cite{tomtwistor} by defining the complex one-forms
\eq{
e^{z^1} = a^{1/2} \left( e^2 + i e^1 \right) \, , \qquad
e^{z^2} = a^{1/2} \left( e^4 + i e^3 \right) \, , \qquad
e^{z^3} = c^{1/2} \left( e^5 + i e^6 \right) \, ,
}
which satisfy
\eq{
d \left( \begin{array}{c} e^{z^1}\\ e^{z^2} \\ e^{z^3} \end{array}\right)=
\left(\begin{array}{cc} - \alpha & 0|_{2 \times 1} \\ 0|_{1 \times 2} & \text{Tr} \, \alpha \end{array}\right)
\left(\begin{array}{c} e^{z^1}\\ e^{z^2} \\ e^{z^3} \end{array}\right)
+\frac{i}{2c^{1/2}}
\left(\begin{array}{c} e^{\bar{z}^2} \wedge e^{\bar{z}^3} \\
e^{\bar{z}^3} \wedge e^{\bar{z}^1} \\
\left(\frac{c}{a}\right) e^{\bar{z}^1} \wedge e^{\bar{z}^2}
\end{array}\right) \, ,
}
with $\alpha$ the anti-hermitian matrix of one forms
\eq{
\alpha = \frac{1}{2} \left( \begin{array}{cc} i(\omega^7+\omega^{10}) & -i\omega^8-\omega^9 \\ -i\omega^8+\omega^9 & i(\omega^7-\omega^{10}) \end{array}\right) \, .
}
These equations take (up to conventions) the form of eq.~(3.10) of \cite{tomtwistor},
with $R=2c^{1/2}$ and $\sigma=c/a$.

\subsection{$\frac{\text{G}_2}{\text{SU(3)}}$}\label{G2qSU3}

\begin{tabular}{|c||c|c|c|}
\hline
\multirow{2}{*}{Betti numbers} & $b_1$ & $b_2$ & $b_3$ \\
\cline{2-4}
& 0 & 0 & 0 \\
\hline
\end{tabular}

The $G_2$ structure constants are given by (see e.g.~\cite{G2SU3}):
\eq{\spl{
& f^1{}_{63}=f^1{}_{45}=f^2{}_{53}=f^2{}_{64} = \frac{1}{\sqrt{3}} \, , \\
& f^7{}_{36}=f^7{}_{45}=f^8{}_{53}=f^8{}_{46}=f^9{}_{56}=f^9{}_{34}=f^{10}{}_{16}=f^{10}{}_{52}\\
& =f^{11}{}_{51} =f^{11}{}_{62}=f^{12}{}_{41}=f^{12}{}_{32}
=f^{13}{}_{31}=f^{13}{}_{24}=\frac{1}{2} \, , \\
& f^{14}{}_{43}=f^{14}{}_{56}=\frac{1}{2\sqrt{3}} \, , \qquad f^{14}{}_{21}=\frac{1}{\sqrt{3}} \, , \\
& f^{i+6}{}_{j+6,k+6} = f_{\text{GM}ijk} \, ,
}}
where $E^{7},\cdots,E^{14}$ generate the su(3) subalgebra, and $f_{\text{GM}ijk}$ are
the Gell-Mann structure constants.

The $G$-invariant two-forms and three-forms are spanned by
\eq{
\{e^{12}-e^{34}+e^{56}\}\, , \qquad \{\rho=e^{245}-e^{135}-e^{146}-e^{236},\hat{\rho}=e^{235}+e^{246}+e^{145}-e^{136}\} \, ,
}
respectively, and there are no invariant one-forms. And again the source (if present) must be proportional to $\Re \Omega$.

The most general solution is then given by
\eq{\spl{
J & = a (e^{12} - e^{34} + e^{56}) \, , \\
\Omega & = d \left[ (e^{245}+e^{146}+e^{135}-e^{236}) + i (e^{145}-e^{246}-e^{235}-e^{136}) \right] \, ,
}}
with $a$, the overall scale, the only free parameter and
\eq{\spl{
 a & > 0 \, ,  \qquad \text{metric positivity} \, , \\
 d^2 & = a^{3}, \qquad \text{normalization of } \Omega \, , \\
 c_1 & := -\frac{3i}{2} {\cal W}_1^-= -\frac{\sqrt{3}a}{d} \, , \\
 {\cal W}_2^- & = 0 \, , \\
 e^{2\Phi} m^2 & = \frac{5}{12} c_1^2  \,  .
}}
We conclude that the only possibility for this coset is the nearly-K\"ahler geometry.

\subsection{$\frac{\text{SU(3)}\times \text{U(1)}}{\text{SU(2)}}$}

\begin{tabular}{|c||c|c|c|}
\hline
\multirow{2}{*}{Betti numbers} & $b_1$ & $b_2$ & $b_3$ \\
\cline{2-4}
& 1 & 0 & 0 \\
\hline
\end{tabular}

The most general case corresponds to taking
\eq{\spl{
 E_{i}&=G_{i+3}, \quad i=1,\dots, 5; \quad E_6=M; \\
E_7&=G_1; \quad E_8=G_2;\quad E_9=G_3
~,}}
where the $G_i$'s are the Gell-Mann matrices generating su(3), $M$ generates a u(1), and
the su(2) subalgebra is generated by $E_7,E_8,E_9$. It follows that
the SU(2) subgroup is embedded entirely inside the SU(3),
so that the total space is given by $\frac{\text{SU(3)}}{\text{SU(2)}} \times \text{U(1)}\simeq S^5\times S^1$.
The structure constants are
\eq{
f^7{}_{89}=1, \quad f^7{}_{14}=f^7{}_{32}=f^8{}_{13}=f^8{}_{24}=f^9{}_{12}=f^9{}_{43}=1/2,
\quad f^{5}{}_{12}=f^{5}{}_{34}=\frac{\sqrt{3}}{2}, \quad\text{cyclic}
~.}
There is a solution for non-zero source:
\eq{\spl{
J & =  -a (e^{13}-e^{24})+b(e^{14}+e^{23})+c e^{56} \, , \\
\Omega & = -\frac{\sqrt{3}}{2c_1} \Big\{\left[ 2a (e^{145}+e^{235}) + 2b (e^{135}-e^{245}) + c(e^{126} + e^{346})\right]  \\
& - \frac{i}{\sqrt{a^2+b^2}} \left[ac(e^{146}+e^{236}) +bc (e^{136}-e^{246}) -2 (a^2+b^2)(e^{125}+e^{345}) \right]
\Big\} \, ,
}}
with $a,b$ and c three free parameters and
\eq{\label{su3qu1su2sol}
\spl{
c & > 0 \, , \quad a^2+b^2 \neq 0 \, , \qquad \text{metric positivity} \, , \\
\frac{1}{(c_1)^2} & = \frac{2}{3}{\sqrt{a^2+b^2}}, \qquad \text{normalization of } \Omega \, , \\
c_1 & := -\frac{3i}{2} {\cal W}_1^- \, , \\
{\cal W}_2^- & = \frac{i}{2\, c_1\sqrt{a^2+b^2}} \left[ -a(e^{13}-e^{24})+b(e^{14}+e^{23}) - 2 c e^{56}\right] \, , \\
d {\cal W}_2^- & = -\frac{i\sqrt{3}}{2\,c_1\sqrt{a^2+b^2}} \left[ a(e^{145}+e^{235}) + b(e^{135}-e^{245}) - c (e^{126}+e^{346})\right] \, , \\
 3|{\cal W}_1^-|^2 - |{\cal W}_2^-|^2 & = 0 \, .
}}
Note that $d {\cal W}_2^-$ is not proportional to $\Re\Omega$, hence
the source is not of the form (\ref{jo}). Interestingly, if we take the part of the source along $\Re \Omega$
to be zero, i.e.\ $j^6 \wedge \Im \Omega=0$,  we find from the last equation in \eqref{su3qu1su2sol}
that $m=0$. This would amount to a combination of smeared D6-branes and O6-planes such that the total tension
is zero. Allowing for negative total tension (more orientifolds), we could have $m > 0$.

\subsection{The remaining cosets}

We now turn to the
remaining cosets of table \ref{taba}. These will be shown to either be equivalent
to one of the previously examined cases, or to support no solution at all. We give
some details in each case for the sake of completeness.

\subsubsection*{$\frac{\text{SU(2)}^2 \times \text{U(1)}}{\text{U(1)}}$}

The most general case corresponds to taking
\eq{\spl{
E_i&=L_i, \quad i=1,2,3; \quad E_{i+3}=L^{\prime}_i, \quad i=1,2; \quad E_6=M; \\
E_7&=L^{\prime}_3-a L_3-bM, \quad a,b\in\mathbb{R}
~,}}
where $\{L_i\}$, $\{L^{\prime}_i\}$ each generates an su(2) algebra, $M$ generates
a u(1) component, and the
u(1) subalgebra is generated by $E_7$.

\begin{tabular}{|c|c||c|c|c|}
\hline
\multirow{4}{*}{Betti numbers} & \multirow{2}{*}{$b=0$} & $b_1$ & $b_2$ & $b_3$ \\
\cline{3-5}
& & 1 & 1 & 2 \\
\cline{2-5}
& \multirow{2}{*}{$b \neq 0$} & $b_1$ & $b_2$ & $b_3$ \\
\cline{3-5}
& & 0 & 0 & 2 \\
\hline
\end{tabular}

For $b \neq 0$, we will show below that the space
(with its SU(3)-structure) is equivalent to the SU(2)$\times$SU(2) example of section \ref{SU2SU2}. For $b=0$, we obtain
the space $T^{1,1}\times \text{U(1)}$ (see e.g.~\cite{t11}) --
which is topologically $S^3 \times S^2 \times S^1$. On this latter space it is possible to find a type IIB SU(2)-structure solution which is T-dual to the solution
on SU(2)$\times$SU(2) of section \ref{SU2SU2}.

The structure constants are then given by
\eq{\spl{
\label{strucSU22U1qU1}
f^{1}{}_{23}=f^{7}{}_{45}=1, \quad \text{cyclic}, \, \\
f^3{}_{45}=f^2{}_{17}=f^1{}_{72}=a, \quad f^6{}_{45}=b~.
}}

There is a nearly-K\"ahler solution for $a=1$ and $b \neq 0$:
\eq{\label{nksolSU22U1qU1}\spl{
J & = k_1 \left[\frac{1}{\sqrt{3}} (e^{15}+e^{24}) + k_3 e^{36} \right] \, , \\
\Omega & = k_2 \left\{ \frac{1}{\sqrt{3}} \left( e^{235}-e^{134}\right)+k_3 \left( e^{126}+e^{456}-b e^{345}\right) \right. \\
& \left. +i \left[ \frac{k_3}{\sqrt{3}} \left( e^{456} -e^{126} + 2 e^{256} - 2 e^{146}\right)
+\frac{1}{3} \left( 2 e^{123}-e^{134}+ e^{235}+e^{345}\right)\right]\right\} \, ,
}}
with
\eq{\spl{
k_3 & = \frac{1}{\sqrt{3} b} \, , \\
k_2^2 & = \frac{2}{3} k_1^3 \qquad \text{normalization of } \Omega \, , \\
c_1:= & -\frac{3i}{2} {\cal W}_1^- = -\frac{k_1}{k_2} \, , \\
{\cal W}_2^- & = 0 \qquad \text{nearly-K\"ahler}.
}}
One can check that the metric is indeed positive definite for all $b \neq 0$. There are also
non nearly-K\"ahler solutions with source not proportional to $\Re \Omega$.

The fact that this coset gives rise to a nearly-K\"{a}hler manifold appears contradictory, as
the list of all such manifolds in six dimensions is exhausted by the examples in sections \ref{SU2SU2}-\ref{G2qSU3} \cite{nkmanifolds}.
The resolution of this puzzle is that the example of the present section is in fact equivalent to the one of
section \ref{SU2SU2}, as we now show.

The structure constants (\ref{strucSU22U1qU1}) correspond to the exterior algebra $d e^I=-1/2f^I{}_{JK}e^{J}\wedge e^{K}$. The latter
is solved explicitly by the following one-forms:\footnote{In the following we set $a=1$ for simplicity; we also set $b=1$,
which can be achieved without loss of generality by a rescaling of the one-form $e^6$.}
\eq{\label{coord}
\spl{
e^1&=\sin\psi_1 d\theta_1-\cos\psi_1 \sin\theta_1 d\phi_1 \, , \\
e^2&=-\cos\psi_1 d\theta_1-\sin\psi_1 \sin\theta_1 d\phi_1 \, ,\\
e^3&=-d\psi_1-d\psi_2-\cos\theta_1 d\phi_1-\cos\theta_2 d\phi_2 \, ,\\
e^4&=\sin\psi_2 d\theta_2-\cos\psi_2 \sin\theta_2 d\phi_2 \, ,\\
e^5&=-\cos\psi_2 d\theta_2-\sin\psi_2 \sin\theta_2 d\phi_2 \, ,\\
e^6&=-d\chi-d\psi_2-\cos\theta_2 d\phi_2 \, ,\\
e^7&=-d\psi_2-\cos\theta_2 d\phi_2~,
}}
where we have introduced the seven coordinates $\chi$, $\phi_{1,2}$, $\theta_{1,2}$, $\psi_{1,2}$.
A straightforward albeit tedious computation reveals that, when expressed in terms of the
coordinates in (\ref{coord}), $J$ and $\Omega$ in \eqref{nksolSU22U1qU1} depend on $\psi_{1,2}$ solely
via the combination $\psi:=\psi_1+\psi_2$. This effectively reduces the coordinate dependence
of $J$ and $\Omega$ to six variables, implying that they indeed parameterize a six-dimensional manifold.
Let us define the one-forms
\eq{
\{g^a\}:=\{e^a\}|_{\psi_1=\psi; ~\psi_2=0}, ~~~~a=1,\dots,6;
}
which manifestly depend on the six coordinates $\chi$, $\psi$, $\phi_{1,2}$, $\theta_{1,2}$. Due to the previous observation, equations
\eqref{nksolSU22U1qU1} still hold if we replace $e^a$ by $g^a$; we will henceforth understand that such a
replacement has been performed.

Let us introduce a new set of one-forms $\hat{g}^a$, defined via:
\al{
\left(\begin{array}{c}
\hat{g}^1\\
\hat{g}^2
\end{array}\right)
=R(-\chi)
\left(\begin{array}{c}
g^1\\
g^2
\end{array}\right); ~~~~~ \hat{g}^3=g^3-g^6; ~~~~~
\left(\begin{array}{c}
\hat{g}^4\\
\hat{g}^5
\end{array}\right)
=R(\chi)
\left(\begin{array}{c}
g^4\\
g^5
\end{array}\right); ~~~~~ \hat{g}^6=g^6
~,}
where
\eq{
R(\chi):=
\left(\begin{array}{cc}
\cos\chi & -\sin\chi\\
\sin\chi & \cos\chi
\end{array}\right)~.
}
It is now
straightforward to check that $J$ and $\Omega$ in (\ref{nksolSU22U1qU1}) can be expressed
solely in terms of the $\hat{g}^a$'s: this can most easily be seen by noting that
\eq{\spl{&\hat{g}^1\wedge\hat{g}^5+\hat{g}^2\wedge\hat{g}^4={g}^1\wedge {g}^5+{g}^2\wedge{g}^4 \, , \\
&\hat{g}^2\wedge\hat{g}^5-\hat{g}^1\wedge\hat{g}^4={g}^2\wedge {g}^5-{g}^1\wedge{g}^4 \, , \\
&\hat{g}^1\wedge\hat{g}^2={g}^1\wedge{g}^2, ~~\hat{g}^4\wedge\hat{g}^5={g}^4\wedge{g}^5~.}}
On the other hand, one can check that the $\hat{g}^a$'s obey the su(2)$\oplus$su(2) algebra, $d\hat{g}^a=-1/2\hat{f}^a{}_{bc}\hat{g}^{b}\wedge \hat{g}^{c}$, as given
by the structure constants $ \hat{f}^1{}_{23} = \hat{f}^4{}_{56}=1$ and cyclic permutations. This concludes the proof
of equivalence to the manifold of section \ref{SU2SU2}.

\subsubsection*{$\frac{\text{SU(3)}\times \text{U(1)}^2}{\text{SU(2)}\times \text{U(1)}}$}


The most general possibility is to take
\eq{\spl{
E_i &= G_{i+3}, \quad i=1,\ldots, 4 \, ; \quad E_5=M, \quad E_6=N ;\\
E_{7}&=G_1,\quad
 E_{8}=G_2,\\
 E_{9}&=G_3,\quad
 E_{10}=G_8-a M, \quad a\in\mathbb{R} \, ,
}}
where the Gell-Mann matrices $G_i$ generate the su(3);  $M$, $N$ generate the two u(1)'s;
the su(2)$\oplus$u(1) subalgebra is generated by $E_{7},\dots, E_{10}$.

\begin{tabular}{|c|c||c|c|c|}
\hline
\multirow{4}{*}{Betti numbers} & \multirow{2}{*}{$a=0$} & $b_1$ & $b_2$ & $b_3$ \\
\cline{3-5}
& & 2 & 2 & 2 \\
\cline{2-5}
& \multirow{2}{*}{$a \neq 0$} & $b_1$ & $b_2$ & $b_3$ \\
\cline{3-5}
& & 1 & 0 & 0 \\
\hline
\end{tabular}

This then leads to the following structure constants
\eq{\spl{
f^7{}_{89}&=1, \quad f^7{}_{14}=f^7{}_{32}=f^8{}_{13}=f^8{}_{24}=f^9{}_{12}=f^9{}_{43}=1/2,
\quad f^{10}{}_{12}=f^{10}{}_{34}=\frac{\sqrt{3}}{2}, \quad \text{cyclic}, \\
f^{5}{}_{12}&=  f^{5}{}_{34}=\frac{a\sqrt{3}}{2} ~.
}}
No solution.

\subsubsection*{SU(2)$\times$U(1)$^3$}

\begin{tabular}{|c||c|c|c|}
\hline
\multirow{2}{*}{Betti numbers} & $b_1$ & $b_2$ & $b_3$ \\
\cline{2-4}
& 3 & 3 & 2\\
\hline
\end{tabular}

No solution.

\subsubsection*{$\frac{\text{SU(2)}^2 \times \text{U(1)}^2}{\text{U(1)}^2}$}

The most general case corresponds to taking
\eq{\spl{
E_i&=L_i, \quad E_{i+2}=L^{\prime}_i, \quad i=1,2; \quad E_5=M,~E_6=N;\\
E_7&=L_3-aM; \quad E_8=L^{\prime}_3-dN, \quad a,d\in\mathbb{R}
~,}}
where $\{L_i\}$, $\{L^{\prime}_i\}$ each generates an su(2) algebra, the $M$, $N$ each generate
a u(1) component, and the
u(1)$\oplus$u(1) subalgebra is generated by $E_7, E_8$.

\begin{tabular}{|c|c||c|c|c|}
\hline
\multirow{6}{*}{Betti numbers} & \multirow{2}{*}{$a=d=0$} & $b_1$ & $b_2$ & $b_3$ \\
\cline{3-5}
& & 2 & 3 & 4 \\
\cline{2-5}
& \multirow{2}{*}{$a \neq 0,d=0$} & $b_1$ & $b_2$ & $b_3$ \\
\cline{3-5}
& & 1 & 1 & 2 \\
\cline{2-5}
& \multirow{2}{*}{$a \neq 0, d \neq 0$} & $b_1$ & $b_2$ & $b_3$ \\
\cline{3-5}
& & 0 & 0 & 2 \\
\hline
\end{tabular}

The structure constants are then given by
\eq{\spl{
f^{7}{}_{12}&=f^{8}{}_{34}=1, \quad \text{cyclic}, \\
f^5{}_{12}&=a,  \quad f^6{}_{34}=d~.
}}
No solution.

\subsubsection*{$\frac{\text{SU(2)}\times \text{U(1)}^4}{\text{U(1)}}$}

The most general possibility consists of taking
\eq{\spl{
E_i&=L_i, ~~i=1,2;\quad E_{i+2}=M_i,~~i=1,\dots, 4;\\
E_7&=L_3-a M_1,~~a\in\mathbb{R}
~,}}
where the $G_i$'s generate the su(2), the $M_i$'s each generate a u(1),
and the u(1) subalgebra is generated by $E_7$.

\begin{tabular}{|c|c||c|c|c|}
\hline
\multirow{4}{*}{Betti numbers} & \multirow{2}{*}{$a=0$} & $b_1$ & $b_2$ & $b_3$ \\
\cline{3-5}
& & 4 & 7 & 8 \\
\cline{2-5}
& \multirow{2}{*}{$a \neq 0$} & $b_1$ & $b_2$ & $b_3$ \\
\cline{3-5}
& & 3 & 3 & 2 \\
\hline
\end{tabular}

The structure constants are
\eq{
f^1{}_{27}=1,\quad \text{cyclic}, \quad f^3{}_{12}=a~.
}
No solution.

\subsubsection*{$\frac{\text{SU(2)}^3}{\text{SU(2)}}$}

\begin{tabular}{|c||c|c|c|}
\hline
\multirow{2}{*}{Betti numbers} & $b_1$ & $b_2$ & $b_3$ \\
\cline{2-4}
& 0 & 0 & 2 \\
\hline
\end{tabular}

The first  possibility corresponds to taking the SU(2) to be diagonally embedded in SU(2)$^3$.
The generators are taken as follows:
\eq{\spl{
E_i&=L_i,\quad E_{i+3}=L^{\prime}_i \, , \\
 E_{i+6}&=L_i+L^{\prime}_i+L^{\prime\prime}_i, ~~i=1,2,3
~,}}
where $\{L_i\}$, $\{L^{\prime}_i\}$, $\{L^{\prime\prime}_i\}$ generate an su(2) each, and the su(2)
subalgebra is generated by $E_7, E_8, E_9$. The structure constants read:
\eq{\spl{
f^1{}_{23}&=f^{4}{}_{56}=f^{7}{}_{89}=1,\quad \text{cyclic},\\
f^6{}_{75}&=-f^5{}_{76}=f^3{}_{72}=-f^2{}_{73}=1,\\
-f^6{}_{84}&=f^4{}_{86}=-f^3{}_{81}=f^1{}_{83}=1,\\
f^5{}_{94}&=-f^4{}_{95}=f^2{}_{91}=-f^1{}_{92}=1~.
}}
Exactly the same nearly-K\"ahler solution as \eqref{nksolSU2SU2}-\eqref{nksolparSU2SU2} is possible.
This coset is equivalent to SU(2)$\times$SU(2).

\begin{tabular}{|c||c|c|c|}
\hline
\multirow{2}{*}{Betti numbers} & $b_1$ & $b_2$ & $b_3$ \\
\cline{2-4}
& 0 & 0 & 2 \\
\hline
\end{tabular}

The other  possibility corresponds to taking the SU(2) to be diagonally embedded in the last two SU(2) factors.
The corresponding generators are taken as follows:
\eq{\spl{
E_i&=L_i,\quad E_{i+3}=L^{\prime}_i \, , \\
 E_{i+6}&=L^{\prime}_i+L^{\prime\prime}_i, ~~i=1,2,3
~,}}
where $\{L_i\}$, $\{L^{\prime}_i\}$, $\{L^{\prime\prime}_i\}$ generate an su(2) each, and the su(2)
subalgebra is generated by $E_7, E_8, E_9$. The structure constants read:
\eq{\spl{
f^1{}_{23}&=f^{4}{}_{56}=f^{7}{}_{89}=1,\quad \text{cyclic},\\
f^6{}_{75}&=-f^5{}_{76}=-f^6{}_{84}=f^4{}_{86}=f^5{}_{94}=-f^4{}_{95}=1~.
}}
No solution.


\subsubsection*{$\frac{\text{SU(2)}^3\times \text{U(1)}}{\text{SU(2)}\times \text{U(1)}}$}

\begin{tabular}{|c||c|c|c|}
\hline
\multirow{2}{*}{Betti numbers} & $b_1$ & $b_2$ & $b_3$ \\
\cline{2-4}
& 0 & 0 & 2 \\
\hline
\end{tabular}

Here we take the SU(2) to be diagonally embedded in the last two SU(2) factors.
The corresponding generators are taken as follows:
\eq{\spl{
E_i&=L_i,\quad E_{i+3}=L^{\prime}_i,\quad E_{10}=L_3+M,\\
 E_{i+6}&=L^{\prime}_i+L^{\prime\prime}_i, ~~i=1,2,3
~,}}
where $\{L_i\}$, $\{L^{\prime}_i\}$, $\{L^{\prime\prime}_i\}$ each generate an su(2), $M$
generates a u(1), and the su(2)$\oplus$u(1)
subalgebra is generated by $E_7, \dots, E_{10}$. The structure constants read:
\eq{\spl{
f^1{}_{23}&=f^{4}{}_{56}=f^{7}{}_{89}=1,\quad \text{cyclic},\\
f^6{}_{75}&=-f^5{}_{76}=-f^6{}_{84}=f^4{}_{86}=f^5{}_{94}=-f^4{}_{95}=f^2{}_{10,1}=-f^1{}_{10,2}=1~.
}}
No solution.

\subsubsection*{$\frac{\text{SU(3)}\times \text{SU(2)}^2}{\text{SU(3)}}$}

This equivalent to the example of section \ref{SU2SU2}.


\section{Interpolations and domain walls}\label{sec:inter}

In this section we put forward a simple ansatz in order to construct supersymmetric  interpolations and supersymmetric domain walls. Our starting point will be the
$\text{AdS}_4$ solutions presented in section \ref{cbc}. We recall that each of these
solutions is of the form:
\al{\label{adssol}
ds^2=ds^2(\text{AdS}_4)+ds^2(\mathcal{M}_6)
~,}
%
where $\mathcal{M}_6$ is a six-dimensional manifold of SU(3)-structure. More specifically,
as reviewed in section \ref{ads4solutions},
its intrinsic torsion is contained in the two torsion classes $\mathcal{W}^-_1$ and $\mathcal{W}_2^-$. In other
words, $\mathcal{M}_6$ is a special case of a {\it half-flat} manifold\footnote{A generic
half-flat manifold has intrinsic torsion contained in $\mathcal{W}_1\oplus\mathcal{W}_2\oplus\mathcal{W}_3$.}.
%
As is well known, six-dimensional half-flat manifolds $\mathcal{M}_6$ lift
via Hitchin flow to seven-dimensional
manifolds $\mathcal{M}_7$ of $G_2$-holonomy \cite{hitc,salamon}:
\al{\label{G2flowmetric}
ds^2(\mathcal{M}_7)=dr^2+g_{mn}(r,y)dy^mdy^n
~,}
where $g_{mn}$ is the $r$-dependent metric of $\mathcal{M}_6$ compatible
with the $r$-dependent solution ($J$,$\Omega$) of the Hitchin-flow equations. This
construction
is reviewed in appendix \ref{sec:hitc}, to which the reader is referred for more details.

\subsubsection*{Interpolations}

In the present context of
supersymmetric solutions to ten-dimensional supergravity,
one would like to construct a physical
realization of the Hitchin flow as follows: we expect that the
AdS$_4\times\mathcal{M}_6$ solutions, presented in section \ref{cbc}, can be
obtained as near-horizon limits of supergravity solutions with brane sources. Assuming this is indeed the case, one would like to construct ten-dimensional supergravity solutions which interpolate between the
`near-horizon' metric (\ref{adssol}) and
\al{
ds^2=ds^2(\mathbb{R}^{1,2})+ds^2(\mathcal{M}_7)
~,}
far from the brane sources, where $ds^2(\mathcal{M}_7)$ is the G$_2$-holonomy metric (\ref{G2flowmetric}).

\subsubsection*{Domain walls}

\begin{figure}
\centering
\psfrag{M4}{$\mathcal{M}_4$}
\psfrag{M6}{$\mathcal{M}_6$}
\psfrag{AdS}{AdS$_4$}
\psfrag{R}{$\mathbb{R}^{1,3}$}
\psfrag{DW}{DW}
\includegraphics[width=8cm]{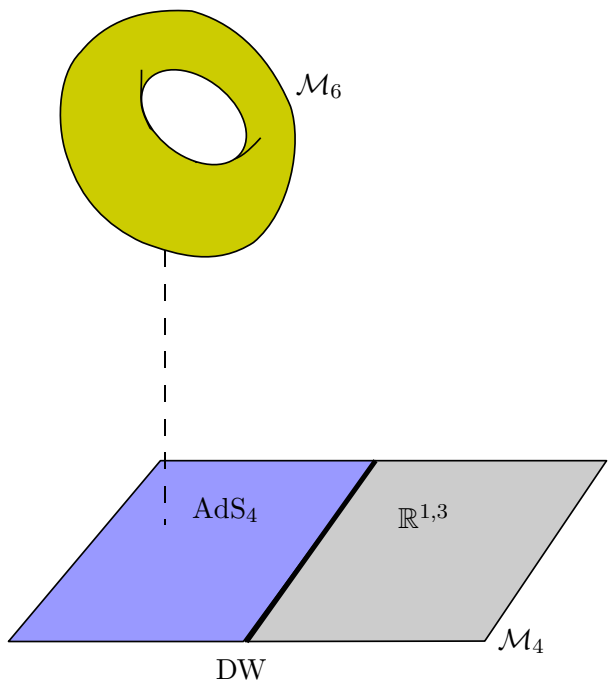}
\caption{A domain wall in four noncompact dimensions $\mathcal{M}_4$ separating
a region of AdS$_4$ from a region of $\mathbb{R}^{1,3}$. The internal manifold
$\mathcal{M}_6$ is fibered over $\mathcal{M}_4$.
Far from the wall $\mathcal{M}_6$ should be independent of $r$, the distance from the wall.}
\label{dwfig}
\end{figure}

Alternatively one could form a (infinitely thin) domain wall in four dimensions, by
patching together
two solutions with different cosmological constants\footnote{see \cite{koun}
for a recent discussion with explicit solutions.} (see figure \ref{dwfig}). The solutions are patched along
a three-dimensional hypersurface (the wall) across which the fluxes, as well as
the first derivative of the metric, are discontinuous. Accordingly,
the wall can be viewed as sourced by localized (infinitely thin) branes.

 To obtain
a solution with a smooth metric, one has to pass from the infinitely-thin wall
approximation to a picture where the wall (and therefore the source branes)
become `thick', i.e.\ acquire a finite extent in the transverse direction.

\subsubsection*{Universal ansatz}

Motivated by the symmetries of the physical problem, we will here
take the metric to be of the form:
\al{\label{s1}
ds^2=e^{2A(r)}\left( ds^2(\mathbb{R}^{1,2}) +dr^2\right) + g_{mn}(r,y)dy^mdy^n
~,}
where $A$ is a real, $r$-dependent function.
Note that any metric of the form
\al{
e^{2U(r)}ds^2(\mathbb{R}^{1,2}) +e^{2V(r)}dr^2
~,}
can be rewritten, by a suitable coordinate transformation $r\rightarrow F(r)$,
as the flat metric of $\mathbb{R}^{1,3}$, up to an $r$-dependent conformal factor and is
thus included in the above ansatz.
To render the problem tractable we will impose a further simplification. Namely
we assume that the internal metric is of the form:
\eq{
\label{metricscale}
g_{mn}(r,y) = \omega^2(r) g_{mn}(y)
~,}
for some $r$-dependent function $\omega$, and we can take $\mathcal{M}_6$
to be any one of the
six-dimensional cosets listed in table \ref{tab2}.
With this metric ansatz we will be able to treat both interpolating
supersymmetric solutions and supersymmetric domain walls simultaneously.
The two cases  differ only in their asymptotics:
\eq{
\mathrm{Interpolation:} \qquad \omega(r)=\left\{\begin{array}{ccl} \mathrm{const}& & r\rightarrow 0\\ \mathrm{const} \times  r &  & r\rightarrow
r_{\infty} \end{array} \right.}
where $r\rightarrow 0$, $r\rightarrow r_{\infty}$
is the near-horizon, far-from-the-source limit respectively,
and
\eq{\label{dwas}
\mathrm{Domain ~Wall:} \qquad \omega(r)= \mathrm{const} \quad r\rightarrow r_{\pm\infty}
}
where  $r\rightarrow r_{\pm\infty}$  is the limit far from the domain wall,
on either side of the wall.
Note that in the case of interpolations, in the $r\rightarrow r_{\infty}$ limit,
the ten-dimensional space-time asymptotes $\mathbb{R}^{1,3}\times \mathcal{M}_7$, where
the metric $ds^2(\mathcal{M}_7)$ (cf.~(\ref{G2flowmetric}))
is a cone over $\mathcal{M}_6$. As
explained in appendix \ref{sec:hitc}, for $\mathcal{M}_7$ to have $G_2$-holonomy
$ds^2(\mathcal{M}_6)$ has to be nearly-K\"{a}hler. This is of course possible
for all six-dimensional cosets listed in table \ref{tab2}.

\subsection{Supersymmetry}\label{sec:susy}

In this section we will formulate and solve the equations following from imposing
the condition of $\mathcal{N}=1$ supersymmetry in three dimensions (two real supercharges).
At the asymptotic limits of the solution, supersymmetry is enhanced to $\mathcal{N}=1$ in four dimensions
(four real supercharges).

Let us now describe the ansatz of the solution. The analysis is a straightforward
 generalization of the calculation in \cite{ltads}, except we use here the conventions of \cite{kt} (see footnote \ref{convfootn}). The spin connection can be read off of eq.~(\ref{s1}):
\eq{\spl{
\nabla_{\mu} & = \partial_{\mu}+\frac{1}{2}A^{\prime}\Gamma_{\mu}\Gamma^r~, \quad \mu=0,1,2~; \qquad \nabla_r=\partial_r;\\
\nabla_m & = {\stackrel{\circ}{\nabla}}{}_m+\frac{1}{4}g^{\prime}_{mn}\Gamma^n\Gamma^r~, \quad m=4,\dots, 9~;
\qquad {\stackrel{\circ}{\nabla}}{}_m:=\partial_m+\frac{1}{4}\omega_{mnl}\Gamma^{nl}
~,}}
where the primes denote differentiation by $r$. In deriving the above we have imposed
the following gauge on the vielbein of the internal metric:
\al{\label{vielgauge}
e^{\prime}_n{}^a=h_n{}^me_m{}^a, ~~~~~h_{mn}:=\frac{1}{2}g^{\prime}_{mn}
~,}
as in \cite{toma}.
Moreover, we will assume that
the $r$-dependence of the internal metric is such that
\eq{
g^{\prime}_{mn}=\frac{2\omega^{\prime}(r)}{\omega(r)}g_{mn}
~,}
for an $r$-dependent function $\omega$. This will be the case if the vielbein is of the form
\al{\label{vielscale}
e_{m}{}^a(r)={\omega(r)}{e}_{m}{}^a(r_0)
~,}
which also automatically satisfies the gauge (\ref{vielgauge}). A priori, the NSNS three-form as well as the
RR forms need only preserve three-dimensional Poincar\'e invariance so they take the form
\eq{\spl{
F_l & = \text{vol}_3 \wedge \left( e^{A} dr \wedge \tilde{F}_{l-4} + \tilde{F}_{\text{3d},l-3} \right) + \hat{F}_l + e^{A} dr \wedge \hat{F}_{r,l-1} \, , \qquad l=0,2,4 \, ;  \\
H & = \hat{H} + H_{\text{3d}} + e^{A} d r \wedge H_{r} \, .
}}
However, we will set
\eq{
\label{formfieldassump}
\tilde{F}_{\text{3d},l}=\hat{F}_{r,l}=H_{\text{3d}}=H_r=0 \, .
}
Note that the domain wall solutions found in \cite{koun} as backgrounds generated by brane configurations (before
their near-horizon limit is taken) satisfy assumption \eqref{formfieldassump}, but not \eqref{metricscale}. Let us nevertheless investigate how far we can get by imposing
both these conditions.

We make the standard SU(3)-structure ansatz for our ten-dimensional spinor
\eq{
\e=(a\zeta_+\otimes\eta_+ +a^*\zeta_-\otimes\eta_-)+(
b^*\zeta_+\otimes\eta_-+b \zeta_-\otimes\eta_+)
\label{spinoransatz}
~,}
where the complex functions $a$, $b$ and the internal unit spinor $\eta$ are a priori
allowed to be $r$-dependent. The ten-dimensional gamma-matrices
decompose correspondingly as:
\eq{\spl{
\G^\mu & = \g^\mu\otimes 1, \qquad \mu=0,\dots, 3;\\
\G^m & = \g_5\otimes \g^m, \qquad  m=4,\dots, 9~.
}}

Furthermore, we impose the following projection on the four-dimensional spinor $\zeta$
\eq{
\label{dwsusy}
\zeta_+=e^{i\theta} e^{-A}\gamma_r\zeta_-
~,}
which reduces, in general, the supersymmetry of the ansatz from four to two real supercharges.
The exponential factor is due to the inverse vielbein, used to convert the curved index on the gamma matrix to a flat one.
In the AdS limit $e^{-i\theta}$ becomes the phase of $W$ defined in \eqref{defW}. In fact, we can always reabsorb this phase
into a redefinition of $\zeta_\pm$ in \eqref{dwsusy} and subsequently in $b/a$ in \eqref{spinoransatz}. Indeed, it will only
ever appear in the combination $e^{i\chi} =(b/a) e^{-i\theta}$.

With the above assumptions, we are ready to proceed to the analysis of the supersymmetry equations, i.e.\ the
vanishing of the gravitino and dilatino variations\footnote{
As an alternative to studying the gravitino and dilatino variations directly,
it is possible to obtain such domain-wall or interpolating solutions as considered here
using the polyform differential equation of appendix A of \cite{km} -- which generalizes
the pure spinor equations for four-dimensional compactifications found in \cite{granaN1}. This approach will be pursued elsewhere \cite{DWfurther}.}.
After a lengthy but straightforward calculation we find that the solution takes
the following form:
\boxedeq{
\label{susysol}
\spl{
e^{i\chi} & =(b/a) e^{-i\theta}=\mathrm{const}; \quad |a|=|b|=\mathrm{const}\times e^{\frac{1}{2}A}; \quad \partial_r \eta_\pm = 0 \, ; \\
H^{(0)} & = -e^{-A}\cos\chi\left( 2 A' - \Phi'+3\frac{\omega'}{\omega} \right) \, , \\
m & = -e^{-A-\Phi} \cos\chi \left( 5 A'- 3 \Phi'+6\frac{\omega'}{\omega} \right) \, , \\
f & = e^{-A-\Phi} \sin\chi \left(  3A'- \Phi'\right) \, , \\
F_2^{(0)} & = e^{-A-\Phi} \sin\chi \left(\frac{1}{3} A'+\frac{1}{3}\Phi'\right) \, , \\
F_4^{(0)} & = - e^{-A-\Phi} \cos\chi \left( 3 A'- \Phi'+ 2 \frac{\omega'}{\omega}\right) \, , \\
\mathcal{W}_1^- & = \frac{2i}{3} H^{(0)} \tan\chi ; \qquad \mathcal{W}_2^- = -i e^{\Phi} F_2'~,
}}
%
where in form notation we have:
\boxedeq{
\label{formform}
\spl{
H&= H^{(0)} \Re\Omega \, , \\
F_{2} & =F_{2}^{(0)} J+ F_2' \, ,\\
F_{4} & = f \text{vol}_4 + \frac{1}{2} F_{4}^{(0)} J\wedge J~.
}}
Note that we are allowing the mass parameter $m$ to be a function of $r$, in order to allow
for the presence of D8-brane sources.

As a consequence of (\ref{vielscale}), $J$ scales as ${\omega^2(r)}$, while $\Omega$ scales as ${\omega^3(r)}$.
Taking the equations (\ref{torsionclasses}) into account, it follows that $\mathcal{W}_1^-$ scales as $1/\omega(r)$,
while $\mathcal{W}_2^-$ scales as ${\omega(r)}$. Comparing
with the last line of (\ref{susysol}), we arrive at the following equations:
\eq{\spl{\label{hcon}
H^{(0)} & = h~\frac{1}{\omega(r)} \, , \\
F_2'&=f_2'~{\omega(r)} e^{-\Phi}
~,}}
where $h$ and $f_2'$ are $r$-independent. {}From (\ref{susysol}), taking (\ref{hcon}) into account, we arrive
at the following constraint:
\boxedeq{
\label{constrsol}
\omega e^{-A}(\frac{\omega^{\prime}}{\omega}+\frac{2}{3}A^{\prime}-\frac{1}{3}\Phi^{\prime})=\mathrm{const}
~,}
where the constant on the right-hand side is equal to $-h/3\cos\chi$.

The AdS$_4$ limit of the above equations  corresponds to $\Phi$, $\omega=$ const, $e^A= R/r$. Indeed
upon setting $\Phi^{\prime}$, $\omega^{\prime}=0$, the reader can verify that eqs.~(\ref{susysol}) reduce precisely
to the solution \eqref{ltsol}, provided we identify:
\eq{
W= e^{-i\theta} A^{\prime}e^{-A} \, .
}
%

{} From the Bianchi identities of the form fields we find that the configuration generically has sources described
by a current $j$ that has an $r$-index. These are indeed domain wall sources. We should still require that
these satisfy appropriate calibration conditions \cite{gencal,km}. It is not so difficult to check that if
\eqref{btildsol} holds, this is automatic for the solution of \eqref{formform}.

{\it In summary}: The solution to the supersymmetry equations is given by eqs.~(\ref{susysol}),
supplemented by the constraint eq.~(\ref{constrsol}), where the form fields are
given by eqs.~(\ref{formform}) and (\ref{hcon}). It is also straightforward to check that requiring that the
Bianchi identities be solved without such sources, reduces to the AdS$_4$ solutions of \cite{ltads}.

\subsection{Explicit profiles}

The solution to the supersymmetry equations of section \ref{sec:susy} does not uniquely specify the profiles
for the warp factors $A$, $\omega$ and the dilaton $\Phi$: given a profile for two of these, the constraint (\ref{constrsol}) can be solved for the third, while (\ref{susysol}) merely solves for all
remaining fields in terms of $A$, $\omega$ and $\Phi$.

\subsubsection*{Interpolations}

Here we will allow for the presence of general (calibrated) sources, so that the sourceless Bianchi identities
(and form-field equations of motion) are violated. The solution to the
supersymmetry equation still allows for considerable freedom in the choice of sources. For concreteness we will
present a specific solution corresponding to constant dilaton and the following profile for the warp factor:
\boxedeq{
\label{aans}
\Phi=\mathrm{const} \, ; \qquad e^A=1+\frac{1}{r}~.
}

Eq.~(\ref{aans}) ensures that the noncompact space interpolates between AdS$_4$ in the $r\rightarrow0$ limit and
$\mathbb{R}^{1,3}$ in the $r\rightarrow\infty$ limit, as follows from the ten-dimensional metric ansatz \eqref{s1}.

We will also assume in addition that the internal
six-dimensional space is nearly-K\"{a}hler, i.e.\ $\mathcal{W}^-_2=0$, so that:
\boxedeq{
\label{btildsol}
F_2'=0~,
}
as follows from (\ref{susysol}). As explained in appendix \ref{sec:hitc}, this allows us to integrate the
Hitchin flow equations, so that the six-dimensional nearly-K\"{a}hler manifold $\mathcal{M}_6$ can be lifted to a seven-dimensional cone $\mathcal{M}_7$ of G$_2$ holonomy:
\al{
ds^2(\mathcal{M}_7)=dr^2 + r^2 g_{mn}(y)dy^mdy^n
~.}
Comparing with (\ref{s1}) and \eqref{metricscale}, we see that in order for the ten-dimensional metric to interpolate between
AdS$_4\times\mathcal{M}_6$ in the $r\rightarrow0$ limit and $\mathbb{R}^{1,2}\times\mathcal{M}_7$ in the $r\rightarrow\infty$ limit, we should impose the following asymptotics on $\omega$:
\al{\label{oasym}
{\omega(r)}=\left\{\begin{array}{ccl} \mathrm{const}& & r\rightarrow 0\\ \mathrm{const} \times r &  & r\rightarrow\infty \end{array} \right.
~~~~~.}
It remains to solve the constraint (\ref{constrsol}). To that end, note that the latter can be rewritten as:
\al{
\omega(r) =-\frac{h}{3\cos\chi}e^{-2A(r)/3+\Phi(r)/3}\int^r ds ~e^{5A(s)/3-\Phi(s)/3}
~.}
Taking (\ref{aans}) into account, this can be integrated to give:
\boxedeq{
\label{osol}
\omega(r)=
-\frac{h}{3\cos\chi}\left\{
1+r-\frac{5}{2}(1+r)^{-2/3}{}_2F_1\left(-\frac{2}{3},-\frac{2}{3};\frac{1}{3};-r \right) \right\}
~,}
where the  integration constant was determined by imposing the asymptotics (\ref{oasym}) for small $r$. The hypergeometric
function on the right-hand side above admits an absolutely convergent Taylor-series expansion
for $r\leq1$ (see e.g.~\cite{gr}, \S\S ~9.100-9.102):
\al{
{}_2F_1\left(-\frac{2}{3},-\frac{2}{3};\frac{1}{3};-r \right)=1-\frac{4}{3}r+\mathcal{O}(r^2)
~.}
To analytically continue to $r>1$ (\cite{gr} \S\S~9.154-9.155) one uses the identity:
\al{
{}_2F_1\left(-\frac{2}{3},-\frac{2}{3};\frac{1}{3};-r \right)=(1+r)^{{5}/{3}}{}_2F_1\left(1,1;\frac{1}{3};-r \right)
~,}
together with the fact that the hypergeometric
function on the right-hand side above admits a series expansion of the form:
\eq{
{}_2F_1\left(1,1;\frac{1}{3};-r \right)\sim\frac{1}{r}\log r+\mathcal{O}(\frac{1}{r^2})
~.}
{}From the above discussion we see that (\ref{osol}) indeed satisfies the asymptotics (\ref{oasym}).
However note that $\omega\rightarrow h/2\cos\chi$ as $r\rightarrow 0$ and
$\omega\rightarrow -rh/3\cos\chi$ as $r\rightarrow \infty$, implying that there is an $r_{\star}\in (0,\infty)$ such that
$\omega(r_{\star})=0$.\footnote{Numerical analysis shows that
$r_{\star}\simeq 0.293$.} These asymptotic values for $\omega(r)$ are in fact valid for any solution for which the
profiles for $A$, $\Phi$ obey the same asymptotics as (\ref{aans}). Also in the case where $h=0$, we see from
(\ref{osol}) that $\omega$ vanishes as $r\rightarrow 0$.
We conclude that: {\em for any profile for $A$, $\Phi$ with the same asymptotics as (\ref{aans}), the
warp factor $\omega(r)$ has a zero at finite radius}.

\begin{figure}
\centering
\psfrag{M7}{$\mathcal{M}_7$}
\psfrag{M6}{$\mathcal{M}_6$}
\psfrag{r}{$r$}
\psfrag{r=0}{$r=0$}
\psfrag{r=r*}{$r=r_{\star}$}
\includegraphics[width=8cm]{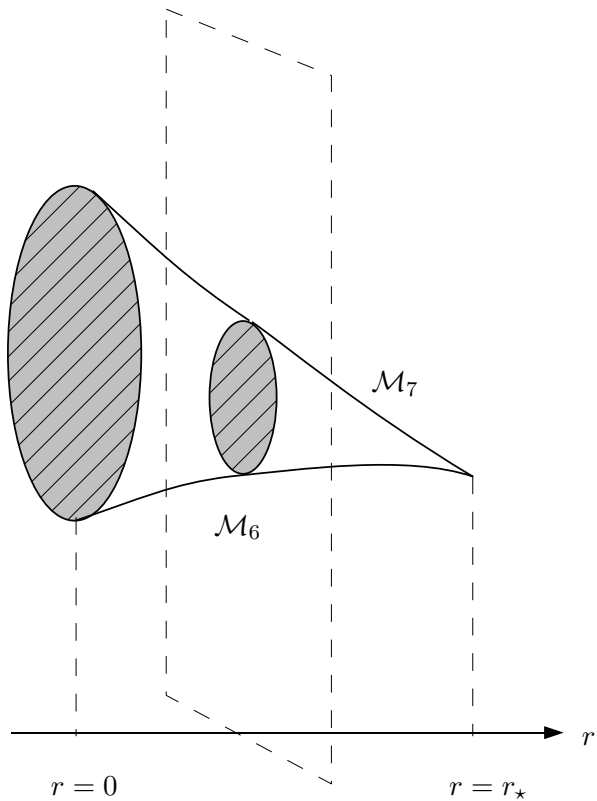}
\caption{A singular  interpolating solution. The internal
six-dimensional manifold $\mathcal{M}_6$ is fibered over
the radial $r$-dimension, forming a seven-dimensional manifold $\mathcal{M}_7$ whose
$r$=constant slices are diffeomorphic to $\mathcal{M}_6$. At $r=r_{\star}$ the
six-dimensional fiber shrinks to zero size.}
\label{singular}
\end{figure}

Plugging eqs.~(\ref{aans}), (\ref{btildsol}), (\ref{osol}) into (\ref{susysol}), allows us to solve for all
remaining fields:
\eq{\spl{\label{above}
f&=-3(1+r)^{-2}\sin\chi\, , \\
m&=5(1+r)^{-2}\cos\chi+\frac{2h}{\omega}\, , \\
F_2^{(0)}&=-\frac{1}{3}(1+r)^{-2}\sin\chi\, , \\
H^{(0)}&=h{\omega}^{-1}\, , \\
F_4^{(0)}&=3(1+r)^{-2}\cos\chi+\frac{2h}{3\omega}~,
}}
where we have set $\Phi=0$ for simplicity. In particular it follows from the
above equations that
the Romans mass blows up in the limit $r\rightarrow r_{\star}$. Moreover, in the
$r\rightarrow \infty$ limit, the NS5 sources also blow up. Indeed,
the profile of the NS5-brane sources can be read off of the Bianchi identity for the three-form:
\al{
dH=j^5
~,}
Using (\ref{above}) and (\ref{formform}) it follows that $j^5$ blows up at infinite radius. We
conclude that the large-radius behaviour of the solution is unphysical (see figure \ref{singular}). Nevertheless, as we will show in
the following, it is possible to obtain a smooth solution interpolating between
AdS$_4$ vacua of different radii.

\section*{Domain walls}

\begin{figure}
\centering
\psfrag{M6}{$\mathcal{M}_6$}
\psfrag{M7}{$\mathcal{M}_7$}
\psfrag{r=0}{$r=-\infty$}
\psfrag{r=infty}{$r=\infty$}
\psfrag{r}{$r$}
\includegraphics[width=8cm]{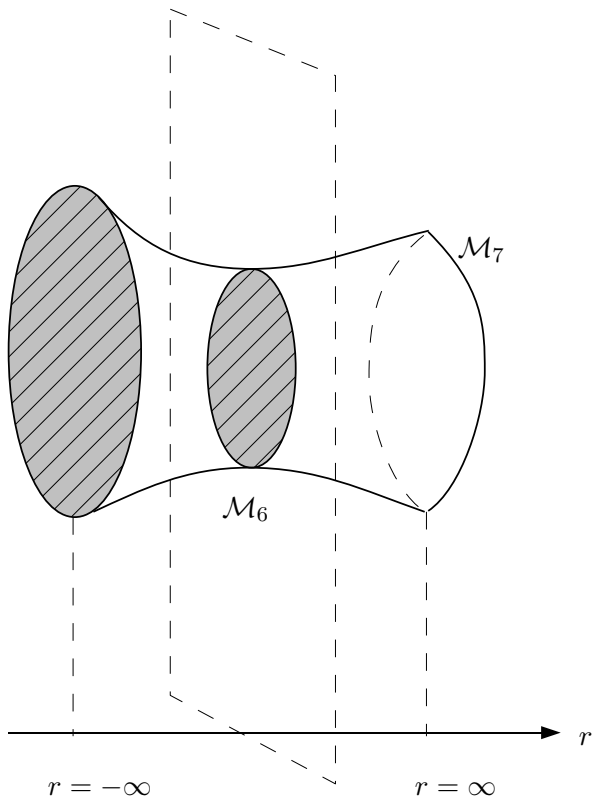}
\caption{A domain wall solution separating two
AdS$_4$ vacua of different radii. The internal
six-dimensional manifold $\mathcal{M}_6$ is fibered over
the radial $r$-dimension, forming a seven-dimensional manifold $\mathcal{M}_7$ whose
$r$=constant slices are diffeomorphic to $\mathcal{M}_6$. As $r\rightarrow \pm\infty$ the
external four-dimensional space asymptotes to AdS$_4$.}
\label{dwfigure}
\end{figure}

Let us now consider a constant dilaton and the following profile for $\omega$:
\boxedeq{
\label{aansadsb}
\Phi=\mathrm{const}; ~~~~~\omega=(2+\tanh r)^{-\frac{2}{5}}
~,}
which satisfies: $\omega\rightarrow\mathrm{const}$, $\omega'\rightarrow0$,
as $r\rightarrow \pm\infty$. Moreover $\omega$ is nowhere-vanishing. The solution
therefore has the appropriate asymptotics \eqref{dwas} for a domain wall. The limits of
$\omega'$ ensure that the domain wall sources, i.e.\ the sources for which $j$ has a component along $r$, vanish
at the endpoints of the radial flow. The constraint (\ref{constrsol}) can be solved
in a closed form to obtain:\footnote{Note that $e^{-A}$ becomes negative
for negative $r$, which amounts to a certain abuse of notation. Equivalently, we could have
introduced a different warp factor: $\Delta:=e^A$,
so that $\Delta$ is well-defined for all $r$.}
\boxedeq{
\label{osoladsb}
e^{-A}=\frac{h}{2\cos\chi}(2+\tanh r)^{-\frac{3}{5}}\left[
2r+\log (\cosh r)\right]
~,}
where we have set the integration constant to zero.
 It can be checked that
$e^{-A}\propto r$,
as $r\rightarrow \pm\infty$, hence the external four-dimensional space asymptotes AdS$_4$
as $r\rightarrow \pm\infty$  (see figure \ref{dwfigure}).

As another example, let us consider again a constant dilaton and the following profile for the warp factor:
\al{
\label{aansads}
\Phi=\mathrm{const}; ~~~~~e^A=\frac{1}{r}+\frac{1}{r+1}
~.}
Eq.~(\ref{aansads}) ensures that the noncompact space interpolates between an AdS$_4$ space
in the $r\rightarrow0$ limit and
another AdS$_4$ space of twice the
radius in the $r\rightarrow\infty$ limit.

Comparing with (\ref{s1}) and \eqref{metricscale}, we see that in order for the ten-dimensional metric to interpolate between two
different space-times of the form
AdS$_4\times\mathcal{M}_6$, we should impose the following asymptotics on $\omega$:
\al{\label{oasymads}
\omega(r)=\mathrm{const}~~~~~r\rightarrow 0,\infty
~.}
As in the previous case, the constraint can be solved for $\omega$ in a closed form:
\al{
\label{osolads}
\omega(r)=
\frac{h}{2\cos\chi}\left\{ 1
-\frac{4r}{1+2r} F_1\left(1,\frac{2}{3},\frac{1}{3};\frac{4}{3};\frac{r}{1+r},\frac{2r}{1+2r}\right)+C\left(r\frac{1+r}{1+2r}\right)^{2/3}
\right\}
~,}
where  the generalized hypergeometric function on the right-hand side above is the
first of Horn's list (sometimes also called the Appell hypergeometric function of two variables), see e.g.~\cite{erde}, \S~5.7.1.
The  integration constant $C$ can be determined by imposing the asymptotics (\ref{oasymads}) for large $r$. Indeed, it can be shown that for large $r$, $F_1\propto r^{2/3}$.

To arrive at
eq.~(\ref{osolads}), we have taken the following identity into account
\al{
-u^2\frac{d}{du}f(u)=\left(u+\frac{u}{1+u}\right)^{\frac{5}{3}}
~,}
where
\al{
f(u):=-\frac{3}{2}\left\{ \left(u+\frac{u}{1+u}\right)^{\frac{2}{3}}
-4u^{-\frac{1}{3}}F_1\left(\frac{1}{3},\frac{2}{3},\frac{1}{3};\frac{4}{3};-\frac{1}{u},-\frac{2}{u}\right)
\right\}
~.}
Furthermore, using identity (1) of \S~5.11 of \cite{erde} we have:
\al{
F_1\left(\frac{1}{3},\frac{2}{3},\frac{1}{3};\frac{4}{3};-\frac{1}{u},-\frac{2}{u}\right)
=u(1+u)^{-\frac{2}{3}}(2+u)^{-\frac{1}{3}}F_1\left(1,\frac{2}{3},\frac{1}{3};\frac{4}{3};\frac{1}{1+u},\frac{2}{2+u}\right)
~.}
Eq.~(\ref{osolads}) then follows from the above upon setting $r=1/u$.

This solution, however, has the problem that the domain wall sources blow up at the endpoints of the radial flow.

\section{Conclusions}

We have reviewed a large class of type IIA
$\mathcal{N}=1$ compactifications to AdS$_4$, based on left-invariant SU(3)-structures
on coset spaces; in the absence of sources they are given in
table \ref{tab2}. The moduli spaces of all solutions contain regions
corresponding to nearly-K\"{a}hler structure, i.e.\ all cosets of
table \ref{tab2} can be viewed as deformations of nearly-K\"{a}hler manifolds,
although in the full quantum theory the `moduli' can only assume discrete values owing to flux quantization.
To our knowledge it is an open question whether or not there exist a manifold with non-vanishing torsion classes $\mathcal{W}^-_{1,2}$ with $d\mathcal{W}^-_{2}\propto \Re\Omega$, such that it cannot be deformed to a nearly-K\"{a}hler manifold.
For that to be the case there would have to be an obstruction to taking the $\mathcal{W}^-_{2}\rightarrow 0$
limit. {}From the physics point-of-view, this would translate to the statement that the
primitive part of the two-form flux has to be non-vanishing.

As we already mentioned in the introduction, the
non nearly-K\"{a}hler deformation of the $\frac{\text{SU(3)}}{\text{U(1)}\times \text{U(1)}}$ coset was recently analysed in \cite{tomtwistor},
using twistor-space techniques. The solution presented here, however,
possesses  one more parameter (for a total of three) in addition to the number of parameters
in the twistor-space construction of \cite{tomtwistor}. The remaining cosets of table \ref{tab2}
have also appeared previously under different guises in the literature, starting with the
early work of Nilsson and Pope on the Hopf-reduction of the (squashed) seven-sphere \cite{np},
and more recently in \cite{font,tomtwistor}.
Here we have put all these cosets in the same context, and have
performed a systematic search for supersymmetric flux compactifications
using the tools of (left-invariant) G-structures.

Allowing for (smeared) six-brane/orientifold sources we obtain more possibilities, listed in table \ref{cab}.
These manifolds can serve as starting points for phenomenologically promising compactifications \cite{acha}.
Given the coset structure of these manifolds, it would certainly be feasible to determine the low-energy physics resulting upon Kaluza-Klein
compactification either in a direct way along the lines of \cite{KK} or using the supersymmetry to make a suitable
ansatz for the expansion forms \cite{palt,effective} and construct the superpotential and K\"ahler potential as in \cite{suppot}.
We leave this interesting line of investigation for future work.

In the last part of the paper we have obtained smooth interpolations between two AdS$_4$ vacua of different radii, using the cosets considered here as internal manifolds.
These solutions can be interpreted as domain walls in the four noncompact dimensions,
and they necessarily contain `thick' branes. However,
we have been unable to obtain physically-sensible profiles
of smooth interpolations between AdS$_4\times \mathcal{M}_6$
and $\mathbb{R}^{1,2}\times \mathcal{M}_7$, where $\mathcal{M}_7$ is the Hitchin lift of $\mathcal{M}_6$. It certainly remains possible that
such profiles do exist for more general ans\"{a}tze than the ones considered here, such as, for example, ans\"{a}tze for which the form fluxes are allowed to have legs in the
radial direction. Another possible generalization is to consider
interpolations where the radial evolution of the internal manifold is not simply given by an overall scaling. We hope to return to this issue in the future.

\begin{acknowledgments}
We would like to thank Luca Martucci, Bengt Nilsson, Paul Smyth and Giorgos Zoupanos for useful discussions.
This work is supported in part by the European Community's Human Potential Programme under contract MRTN-CT-2004-005104 ``Constituents,
fundamental forces and symmetries of the universe'' and the Excellence Cluster ``The Origin and the Structure of the Universe'' in Munich.
P.~K.\ is supported by the German Research Foundation (DFG) within the Emmy-Noether-Program (Grant number ZA 279/1-2).
\end{acknowledgments}

\appendix

\section{The structure group of coset spaces}\label{structuregroup}

In this section we review in some detail the statement that the
tangent bundle of the manifold $G/H$ has structure group $H$.

Let $G$ be a Lie group and $H$ a closed subgroup of $G$.
The group $G$ can be regarded as a  principal bundle, denoted by $G(G/H,H)$, with base
$M=G/H$ and fibre $H$. Moreover, the structure group of $G(G/H,H)$ is $H$  (\cite{koba}, p.~55).
The action of $G$ on $M$ induces a map $f:G(G/H,H)\rightarrow L(M,S)$, where $L(M,S)$
is the frame bundle of $M$ with structure group
$S\subseteq GL(d,\mathbb{R})$, $d:=\mathrm{dim}(M)$, and a corresponding map $\varphi:H\rightarrow S$.
If the action of $G$ is effective (or, equivalently,  $H$ contains no nontrivial
invariant subgroup of $G$), both $f$ and $\varphi$ are
isomorphisms (\cite{koba}, pp.~301-302).\footnote{It is interesting
to note that a similar statement can be formulated for super-coset manifolds: let
$G$ be a super-Lie group and $H$ a closed subgroup of $G$. Then the frame bundle over
$G/H$ is equivalent to the principal bundle $G(G/H,H)$
(\cite{klep}, section 7.2).}

On the other hand, the frame bundle
 $L(M)$ can be regarded as the
associated principal bundle of the tangent bundle of $M$, $T(M)$. In particular,
 $T(M)$ and $L(M)$ have the same structure group (\cite{stee}, pp.~35-36). We conclude, from the discussion in the
preceding paragraph, that the structure group of the tangent bundle of $M=G/H$ is isomorphic to $H$. Note, as
a corollary, that by taking $H=\{e\}$ to consist of the identity element of $G$, it follows that the structure
group of the tangent bundle of $G$, regarded as a manifold, is trivial and the manifold is parallelizable. To summarize:

\medskip
{\it Let $G$ be a Lie group and $H$ a closed subgroup of $G$, such that $H$ contains no nontrivial
invariant subgroup of $G$.
The structure group of the tangent bundle of $M=G/H$ is $H$. }

\medskip
It follows from the above that, as already mentioned in the introduction,
it suffices to take $H$ to be isomorphic to SU(3) or a subgroup thereof.
It then follows that all
possible six-dimensional manifolds $M$ of this type consist
of the ones listed in table \ref{taba}.

Essentially the same list has appeared, for different reasons, in \cite{govi}.
We would like, however, to make a remark about the entries with $H$=S(U(2)$\times$U(1)). It is often incorrectly stated in the physics literature, that SU(2)$\times$U(1) is a subgroup of SU(3). To see why this is inaccurate,
we first quote the following uniqueness theorem (see e.g.~\cite{helgason}, p.~102):

\medskip
{\it Let $G$ be a Lie group with Lie algebra $\mathfrak{g}$, and $\mathfrak{h}$ a subalgebra of $\mathfrak{g}$.
There exists a unique connected Lie subgroup $H$ of $G$, whose Lie algebra is $\mathfrak{h}$. }

\medskip
We will now show that S(U(2)$\times$U(1)) is a subgroup of SU(3), with Lie algebra
$\mathfrak{su}(2)\oplus\mathfrak{u}(1)$. First note that S(U(2)$\times$U(1)) is given by
\al{
\left\{  \left(\begin{array}{cc}
e^{i\phi}& 0\\
0&A
\end{array}\right)~, ~~\mathrm{such ~that:} ~A\in \mathrm{U}(2),~
e^{i\phi}\in \mathrm{U}(1), ~e^{i\phi}\mathrm{det}A=1~ \right\}~,
}
and is clearly a subgroup of SU(3). It is also connected, since it is
isomorphic to U(2).\footnote{
More generally, one can make the identification S(U(n)$\times$U(1))$\cong$U(n), upon which one obtains the
well-known result that $\mathbb{CP}^n\cong$SU(n+1)/U(n) (see e.g.~\cite{wallach}, p.~146).} Indeed, by setting $e^{i\phi}=(\mathrm{det}A)^{-1}$, taking into account
the fact that $|\mathrm{det}A|=1$ for any unitary matrix $A$, we can therefore identify
\al{
\mathrm{S(U(2)}\times \mathrm{U(1))} \cong  \left\{  \left(\begin{array}{cc}
(\mathrm{det}A)^{-1}& 0\\
0&A
\end{array}\right)~, ~~\mathrm{such ~that:} ~A\in \mathrm{U}(2) \right\}\cong\mathrm{U}(2)~.
}
Moreover, it is well-known that
\al{
\mathrm{U}(n)\cong\frac{\mathrm{SU(n)}\times \mathrm{U(1)}}{\mathbb{Z}_n}~.
}
It follows that S(U(2)$\times$U(1))$\cong$U(2) and SU(2)$\times$U(1) are distinct Lie groups,
however they share
the same Lie algebra: $\mathfrak{su}(2)\oplus\mathfrak{u}(1)$.
{} From the uniqueness theorem quoted above, it follows that it is
S(U(2)$\times$U(1)), but not SU(2)$\times$U(1), that is a subgroup of SU(3).

\section{Hitchin flow}\label{sec:hitc}

Six-dimensional half-flat manifolds lift to seven-dimensional
manifolds of $G_2$-holonomy, as follows \cite{hitc}: consider
$\mathcal{M}_7=\mathcal{M}_6\times I$, where $\mathcal{M}_6$ is a six-dimensional
half-flat manifold, and $I$ is an interval parameterized by the
coordinate $r$. Moreover, consider the real three-form $\phi$ defined by:
\al{\label{phid}
\phi=J\wedge dr+\Re\Omega~,
}
where the SU(3)-structure ($J$,$\Omega$) of $\mathcal{M}_6$ is now $r$-dependent. This defines a G$_2$
structure on $\mathcal{M}_7$. The additional requirement that $\mathcal{M}_7$ have G$_2$-holonomy is equivalent
to the requirement that $\phi$ be closed and coclosed. This is, in its turn, equivalent
to the `Hitchin flow' equations \cite{hitc}:
\eq{\spl{\label{hitf}
0&=\hat{d}J-\partial_r \Re\Omega \, ,\\
0&=\hat{d}\Im\Omega-J\wedge\partial_r J~,
}}
where $\hat{d}$ is the restriction of the exterior derivative to $\mathcal{M}_6$.

The metric of $\mathcal{M}_7$ is determined by its G$_2$ structure as follows (see e.g.~\cite{kari}):
define the symmetric two-tensor
\al{
B_{mn} du^1 \wedge \cdots \wedge du^7=(\frac{\partial}{\partial u^m}\lrcorner~\phi)\wedge (\frac{\partial}{\partial u^n}\lrcorner~\phi)\wedge\phi
~,}
where the $u^m$, $m=1,\dots, 7$, are local coordinates on $\mathcal{M}_7$. The metric is then given by:
\al{\label{g2metr}
g_{mn}=\frac{B_{mn}}{6^{\frac{2}{9}}\mathrm{det}(B)^{\frac{1}{9}}}
~.}
{} From (\ref{phid}) and (\ref{g2metr}) we can read off the metric on $\mathcal{M}_7=\mathcal{M}_6\times I$:
\al{
ds^2(\mathcal{M}_7)=dr^2+g_{mn}(r,y)dy^mdy^n
~,}
where $g_{mn}$ is the $r$-dependent metric of $\mathcal{M}_6$ compatible
with the $r$-dependent solution ($J$,$\Omega$) of the Hitchin-flow equations (\ref{hitf}).

The Hitchin flow equations can readily be integrated in the case where the six-dimensional
manifold space is  $\mathcal{M}_6$ is nearly-K\"{a}hler, i.e.\ $\mathcal{W}_2^-=0$. In this case
it can be seen that $\mathcal{M}_7$ is simply a cone over a base $\mathcal{M}_6$. In other
words:
\al{
g_{mn}(r,y)dy^idy^j=r^2g_{mn}(r_0,y)dy^idy^j
~,}
where $r_0$ is some fixed value of the radial coordinate.

%
%


\begin{thebibliography}{99}


\bibitem{gstruc}
  J.~P.~Gauntlett, D.~Martelli, S.~Pakis and D.~Waldram,
  ``G-structures and wrapped NS5-branes,''
  \cmp{247}{2004}{421} [arXiv:\hepth{0205050}];
  G.~L.~Cardoso, G.~Curio, G.~Dall'Agata, D.~L\"ust, P.~Manousselis and G.~Zoupanos,
  ``Non-K\"ahler string backgrounds and their five torsion classes,''
  \npb{652}{2003}{5} [arXiv:\hepth{0211118}];
  J.~P.~Gauntlett, D.~Martelli and D.~Waldram,
  ``Superstrings with intrinsic torsion,''
  \prd{69}{2004}{086002}
  [arXiv:\hepth{0302158}].

\bibitem{getala}
  M.~Gra\~{n}a, R.~Minasian, M.~Petrini and A.~Tomasiello,
  ``Supersymmetric backgrounds from generalized Calabi-Yau manifolds,''
  \jhep{0408}{2004}{046} [arXiv:\hepth{0406137}].

\bibitem{granareview}
  M.~Gra\~{n}a,
  ``Flux compactifications in string theory: A comprehensive review,''
  \prep{423}{2006}{91} [arXiv:\hepth{0509003}].

\bibitem{gengeom}
  N.~Hitchin,
  ``Generalized Calabi-Yau manifolds,''
  Quart.\ J.\ Math.\ Oxford Ser. {\bf 54} (2003) 281
  [arXiv:\Math{DG}{0209099}];
  M.~Gualtieri,
  ``Generalized complex geometry,''
  arXiv:\Math{DG}{0401221}.

\bibitem{granaN1}
  M.~Gra\~{n}a, R.~Minasian, M.~Petrini and A.~Tomasiello,
  ``Generalized structures of $N=1$ vacua,''
  \jhep{0511}{2005}{020} [arXiv:hep-th/0505212].

\bibitem{granascan}
  M.~Gra\~{n}a, R.~Minasian, M.~Petrini and A.~Tomasiello,
  ``A scan for new ${\cal N}=1$ vacua on twisted tori,''
  \jhep{0705}{2007}{031}
  [arXiv:hep-th/0609124].

\bibitem{nogo}
  B.~de Wit, D.~J.~Smit and N.~D.~Hari Dass,
  ``Residual supersymmetry of compactified $D=10$ supergravity,''
  \npb{283}{1987}{165};
  J.~M.~Maldacena and C.~N\'u\~{n}ez,
  ``Supergravity description of field theories on curved manifolds and a no go theorem,''
  \ijmpa{16}{2001}{822} [arXiv:\hepth{0007018}].

\bibitem{behr}
  K.~Behrndt and M.~Cveti\v{c},
  ``General $\mathcal{N} = 1$ supersymmetric flux vacua of (massive) type IIA string theory'',
  \prl{95}{2005}{021601} [arXiv:\hepth{0403049}];
  K.~Behrndt and M.~Cveti\v{c},
  ``General $\mathcal{N} = 1$ supersymmetric fluxes in massive type IIA string theory'',
  \npb{708}{2005}{45} [arXiv:\hepth{0407263}].

\bibitem{ltads}
  D.~L\"{u}st and D.~Tsimpis,
  ``Supersymmetric AdS$_4$ compactifications of IIA supergravity,''
  \jhep{0502}{2005}{027} [arXiv:\hepth{0412250}].

\bibitem{font}
  G.~Aldazabal and A.~Font,
  ``A second look at ${\cal N}=1$ supersymmetric AdS$_4$ vacua of type IIA supergravity,''
  \jhep{0802}{2008}{086} [\arXividhepth{0712.1021}].

\bibitem{np}
  B.~E.~W.~Nilsson and C.~N.~Pope,
  ``Hopf fibration of eleven-dimensional supergravity,''
  \cqg{1}{1984}{499}.

\bibitem{tomtwistor}
  A.~Tomasiello,
  ``New string vacua from twistor spaces,''
  \arXividhepth{0712.1396}.

\bibitem{palt}
  T.~House and E.~Palti,
  ``Effective action of (massive) IIA on manifolds with SU(3) structure,''
  \prd{72}{2005}{026004} [arXiv:\hepth{0505177}].

\bibitem{zoup}
  D.~Kapetanakis and G.~Zoupanos,
  ``Coset space dimensional reduction of gauge theories,''
  \prep{219}{1992}{1}.

\bibitem{cosetheterotic}
  D.~L\"ust,
  ``Compactification of ten-dimensional superstring theories over Ricci-flat
  coset spaces,''
  \npb{276}{1986}{220};
  L.~Castellani and D.~L\"ust,
  ``Superstring compactification on homogeneous coset spaces with torsion,''
  \npb{296}{1988}{143};
  T.~R.~Govindarajan, A.~S.~Joshipura, S.~D.~Rindani and U.~Sarkar,
  ``Supersymmetric compactification of the heterotic string on coset spaces,''
  \prl{57}{1986}{2489}.

\bibitem{govi}
  T.~R.~Govindarajan, A.~S.~Joshipura, S.~D.~Rindani and U.~Sarkar,
  ``Coset spaces as alternatives to Calabi-Yau spaces in the presence of
  gaugino condensation,''
  \ijmpa{2}{1987}{797}.

\bibitem{mpz}
   P.~Manousselis, N.~Prezas and G.~Zoupanos,
   ``Supersymmetric compactifications of heterotic strings with fluxes and
   condensates,''
   \npb{739}{85}{2006}
   [arXiv:\hepth{0511122}].

\bibitem{nkmanifolds}
  J.-B.~Butruille, ``Homogeneous nearly K\"ahler manifolds,''
  arXiv:\Math{DG}{0612655}.

\bibitem{kt}
  P.~Koerber and D.~Tsimpis,
  ``Supersymmetric sources, integrability and generalized-structure
  compactifications,''
  \jhep{0708}{2007}{082}
  [\arXividhepth{0706.1244}].

\bibitem{bovy}
  J.~Bovy, D.~L\"{u}st and D.~Tsimpis,
  ``N = 1,2 supersymmetric vacua of IIA supergravity and SU(2) structures,''
  \jhep{0508}{2005}{056}   [arXiv:\hepth{0506160}].

\bibitem{roma}
  L.~J.~Romans,
  ``Massive $N=2$a supergravity in ten dimensions,''
  \plb{169}{1986}{374}.

\bibitem{democratic}
  E.~Bergshoeff, R.~Kallosh, T.~Ort\'{\i}n, D.~Roest and A.~Van Proeyen,
  {\em New formulations of $D = 10$ supersymmetry and D8 - O8 domain walls},
  \cqg{18}{2001}{3359}
  [arXiv:\hepth{0103233}].

\bibitem{gencal}
  P.~Koerber,
  ``Stable D-branes, calibrations and generalized Calabi-Yau geometry,''
  \jhep{0508}{2005}{099}
  [arXiv:\hepth{0506154}];
  L.~Martucci and P.~Smyth,
  ``Supersymmetric D-branes and calibrations on general $\mathcal{N}$ = 1 backgrounds,''
  \jhep{0511}{2005}{048}
  [arXiv:\hepth{0507099}];
   L.~Martucci,
   ``D-branes on general $\mathcal{N}=1$ backgrounds: Superpotentials and D-terms,''
   \jhep{0606}{2006}{033} [arXiv:\hepth{0602129}].

\bibitem{ktu}
  P.~Koerber and D.~Tsimpis, unpublished.


\bibitem{acha}
  B.~S.~Acharya, F.~Benini and R.~Valandro,
  ``Fixing moduli in exact type IIA flux vacua,''
  \jhep{0702}{2007}{018}
  [arXiv:\hepth{0607223}].

\bibitem{cosetrev1}
  P.~van Nieuwenhuizen,
  ``General theory of coset manifolds and antisymmetric tensors applied to
  Kaluza-Klein supergravity,''
  ITP-SB-84-57, ``Supersymmetry and supergravity '84,'' World Scientific, Singapore (1985).

\bibitem{cosetrev2}
  F.~M\"uller-Hoissen and R.~St\"uckl,
  ``Coset spaces and ten-dimensional unified theories,''
  \cqg{5}{1988}{27}.

\bibitem{knb}
 S.~Kobayashi and K.~Nomizu,
  ``Foundations of Differential Geometry,''
  Vol.~2, Wiley Classics Library Edition, U.S.A.\ (1996).

\bibitem{G2SU3}
  M.~Gunaydin and F.~Gursey,
  ``Quark structure and octonions,''
  \jmp{14}{1973}{1651}; D.~L\"{u}st,
  ``Compactification of ten-dimensional superstring theories over Ricci flat
  coset spaces,''
  \npb{276}{1986}{220}.

\bibitem{t11}
  D.~N.~Page and C.~N.~Pope,
  ``Which compactifications of $D = 11$ supergravity are stable?,''
  \plb{144}{1984}{346}.

\bibitem{hitc}
  N.~Hitchin,
 ``Stable forms and special metrics,''
  arXiv:\Math{DG}{0107101}.

\bibitem{salamon}
  S.~Chiossi and S.~Salamon,
  ``The intrinsic torsion of SU(3) and G$_2$ structures,''
  Differential Geometry, Valencia 2001, World Sci. Publishing, 2002, pp.~115-133
  [arXiv:\Math{DG}{0202282}].

\bibitem{koun}
  C.~Kounnas, D.~L\"ust, P.~M.~Petropoulos and D.~Tsimpis,
  ``AdS$_4$ flux vacua in type II superstrings and their domain-wall solutions,''
  \jhep{0709}{2007}{051}
  [\arXividhepth{0707.4270}].

\bibitem{toma}
  J.~P.~Hsu, A.~Maloney and A.~Tomasiello,
  ``Black hole attractors and pure spinors,''
  \jhep{0609}{2006}{048}
  [arXiv:\hepth{0602142}].

\bibitem{km}
  P.~Koerber and L.~Martucci,
  ``D-branes on AdS flux compactifications,''
  \jhep{0801}{2008}{047}
  [\arXividhepth{0710.5530}].

\bibitem{DWfurther}
  P.~Koerber, L.~Martucci and P.~Smyth, work in progress.

\bibitem{gr}
  I.~S.~Gradsbteyn and L.~M.~Ryzbik,
  ``Table of integrals, series and products,''
  Elsevier Pte Ltd, Singapore (2004).


\bibitem{erde}
  A.~Erd\'elyi ed.,
  ``Higher trancendental functions,''
  Vol.~1, McGraw-Hill Book Company, U.S.A.\ (1953).


\bibitem{KK}
  M.~J.~Duff, B.~E.~W.~Nilsson and C.~N.~Pope,
  ``Kaluza-Klein Supergravity,''
  \prep{130}{1986}{1}.

\bibitem{effective}
  S.~Gurrieri, J.~Louis, A.~Micu and D.~Waldram,
  ``Mirror symmetry in generalized Calabi-Yau compactifications,''
  \npb{654}{2003}{61}
  [arXiv:\hepth{0211102}];
  T.~W.~Grimm and J.~Louis,
  ``The effective action of type IIA Calabi-Yau orientifolds,''
  \npb{718}{2005}{153} [arXiv:\hepth{0412277}];
  A.-K.~Kashani-Poor and R.~Minasian,
  ``Towards reduction of type II theories on SU(3) structure manifolds,''
  \jhep{0703}{2007}{109}
  [arXiv:\hepth{0611106}];
  A.-K.~Kashani-Poor,
  ``Nearly K\"ahler reduction,''
  \jhep{0711}{2007}{026}
  [\arXividhepth{0709.4482}];
  D.~Cassani,
  ``Reducing democratic type II supergravity on SU(3)$\times$SU(3) structures,''
  \arXividhepth{0804.0595}.

\bibitem{suppot}
  M.~Gra\~{n}a, J.~Louis and D.~Waldram,
  ``Hitchin functionals in $N = 2$ supergravity,''
  \jhep{0601}{2006}{008} [arXiv:\hepth{0505264}];
  I.~Benmachiche and T.~W.~Grimm,
  ``Generalized $N = 1$ orientifold compactifications and the Hitchin
  functionals,''
  \npb{748}{2006}{200} [arXiv:\hepth{0602241}];
  P.~Koerber and L.~Martucci,
  ``From ten to four and back again: how to generalize the geometry,''
  \jhep{0708}{2007}{059} [\arXividhepth{0707.1038}];
  P.~Koerber and L.~Martucci,
  ``Warped generalized geometry compactifications, effective theories and
  non-perturbative effects,''
  \arXividhepth{0803.3149}.

\bibitem{koba}
 S.~Kobayashi and K.~Nomizu,
  ``Foundations of Differential Geometry,''
  Vol.~1, Wiley Classics Library Edition, U.S.A.\ (1996).

\bibitem{klep}
  A.~F.~Kleppe and C.~Wainwright,
  ``Super coset space geometry,''
  \jmp{48}{2007}{053511} [arXiv:\hepth{0610039}].

\bibitem{stee}
 N.~Steenrod,
  ``The topology of fibre bundles,''
  Princeton University Press, U.S.A.\ (1951).

\bibitem{helgason}
 S.~Helgason,
  ``Differential geometry and symmetric spaces,'' Academic Press, U.K.\ (1962).

\bibitem{wallach}
 N.~R.~Wallach,
  ``Harmonic analysis on homogeneous spaces,'' Marcel Dekker, U.S.A.\ (1973).

\bibitem{kari}
  S.~Karigiannis,
  ``Geometric Flows on Manifolds with G$_2$ Structure, I,''
  arXiv:\Math{DG}{0702077}.

\end{thebibliography}
\end{document}